\documentclass[eqsecnum,aps,showpacs,floatfix,superscriptaddress,nofootinbib,twocolumn]{revtex4}
\usepackage{amsmath,amssymb,latexsym}
\usepackage[pdftex]{graphicx}

\begin{document}
\title{Strong-field tidal distortions of rotating black
  holes:\\ II.\ Horizon dynamics from eccentric and inclined orbits}
\author{Stephen O'Sullivan}
\affiliation{Department of Physics and MIT Kavli Institute,
  Massachusetts Institute of Technology, Cambridge, MA 02139}
\author{Scott A.\ Hughes}
\affiliation{Department of Physics and MIT Kavli Institute,
  Massachusetts Institute of Technology, Cambridge, MA 02139}
\begin{abstract}
In a previous paper, we developed tools for studying the horizon
geometry of a Kerr black hole that is tidally distorted by a binary
companion using techniques that require large mass ratios but can be
applied to any bound orbit and allow for arbitrary black hole spin.
We now apply these tools to generic Kerr black hole orbits.  This
allows us to investigate horizon dynamics: the tidal field perturbing
the horizon's geometry varies over a generic orbit, with significant
variations for eccentric orbits.  Many of the features of the
horizon's behavior found previously carry over to the dynamical case
in a natural way.  In particular, we find significant offsets between
the applied tide and the horizon's response.  This leads to bulging in
the horizon's geometry which can lag or lead the orbit, depending upon
the hole's rotation and the orbit's geometry.  An interesting and
apparently new feature we find are small-amplitude, high-frequency
oscillations in the horizon's response.  We have not been able to
identify a mechanism for producing these oscillations, but find that
they appear most clearly when rapidly rotating black holes are
distorted by very strong-field orbits.
\end{abstract}


\maketitle

\section{Introduction}
\label{sec:intro}

The study of relativistic tidal deformations and their impact on the
dynamics of compact binaries has received a great deal of attention in
recent years.  Much of this recent activity was kicked off by studies
of tides in systems containing neutron stars {\cite{rmsucf09, hllr10,
    dnv12, dn09, bp09, bdgnr10, wab13, vzh14}}.  Older work had
already demonstrated that tidal coupling was quite important in
systems containing black holes, but used language that clouded the
role of tides, using instead a dual description of tidal coupling as
``radiation down the event horizon'' {\cite{t73, tp74, h00, h01}}.
Recent papers focusing on black holes in binaries have examined in
detail how tides distort black holes and their near-hole geometry.
Most of these papers have focused on non-rotating {\cite{bp09, dl09,
    tp08, pv10, vpm11}} and slowly rotating {\cite{fl05, p15, pgmf15}}
black holes (with Ref.\ {\cite{pgmf15}} discussing tidal distortions
of a broad class of spinning objects).

Our contribution to this body of work has been to develop numerical
tools for characterizing tidally distorted black holes which are good
for strong-field orbits and arbitrary black hole spins.  These tools
are based on black hole perturbation theory, and so assume binaries of
extreme mass ratio: the mass $\mu$ of the small body which is the
source of the tide is much less than the mass $M$ of the black hole
that is tidally distorted.  In Ref.\ {\cite{paperI}} (hereafter
``paper I''), we developed tools for characterizing the tidal field
that acts on a Kerr black hole.  The tools are designed in order to
adapt pre-existing codes which have been used to study gravitational
wave emission from extreme mass ratio binaries (e.g., {\cite{dh06}}).
We also developed tools to visualize a tidally distorted black hole by
embedding the two-dimensional horizon at each moment in some time
slicing in a flat three-dimensional Euclidean space.  These embeddings
are only good for Kerr spin parameter $a/M \le \sqrt{3}/2$; for higher
spins, the horizon cannot be globally embedded in a Euclidean space
even in the absence of a distorting tide {\cite{smarr}}.

Although the tools we developed in paper I are generic and can be
applied to any bound Kerr black hole orbit, we only showed results for
tidal distortions arising from circular and equatorial orbits.  By
focusing on this relatively simple case, we were able to examine some
of the key aspects of event horizon physics in a particularly clean
limit.  For example, paper I examined in some detail the phase offset
between the angle at which the horizon is maximally distorted (the
location of its ``tidal bulge'') and the position of the orbit.  As
has been amply discussed in past literature {\cite{hh72, h73, h74,
    membrane}}, the event horizon acts in many ways like the surface
of a gravitating fluid body; a very readable summary discussion of
this connection can also be found in Cardoso and Pani {\cite{cp2012}}.
The horizon is deformed by tidal stresses, tending to bulge toward the
``moon'' which is the source of the tide.  The bulging response is,
however, not synchronous with the applied tide.  For a fluid body,
viscosity causes the fluid's response to lag the applied tide.  As a
consequence, if the moon's orbit is faster than the body's spin, the
bulge lags the orbit's position.  Conversely, if the orbit is slower
than the spin, then the bulge leads the orbit's position.

At least for very slowly varying tidal fields, this picture describes
the geometry of the black hole's tidal bulge with respect to the orbit
--- provided we swap ``lead'' and ``lag.''  Tides from a moon which
orbits faster than the hole's spin raise a bulge which {\it leads} the
orbit's position; tides from a moon which orbits more slowly {\it lag}
the orbit's position.  The swap of ``lead'' and ``lag'' as compared to
the fluid star is due to the {\it teleological} nature of the event
horizon: how the horizon depends at some moment in a given time
slicing depends upon the stresses that it will feel in the future.
Though this counterintuitive behavior might seem to violate causality,
it is a simple consequence of how the horizon is defined: whether an
event is inside or outside the horizon depends on that event's future.
See paper I and references therein (as well as the references cited
above) for much more detailed discussion of the horizon's teleological
nature and its consequences.

Although circular and equatorial orbits were useful for testing our
tidal distortion toolkit, this limit does not show the full range of
horizon dynamics that can be expected from tidal interactions.
Indeed, the horizon's distortion is stationary in this case, showing
no variation at all in a frame that co-rotates with the orbit.  The
purpose of this paper is to go beyond this limit and to explore how
the horizon responds to generic --- inclined and eccentric --- orbits.
Generic orbits and the tides they produce are dynamical even when
examined in a frame that corotates at the orbit's axial frequency.
Eccentricity is particularly important: at leading order the tidal
field varies as $1/r^3$, so as the orbit's radius varies from $r_{\rm
  max}$ to $r_{\rm min}$, the tidal field varies by a factor $r_{\rm
  max}^3/r_{\rm min}^3$.  If the large black hole spins, even constant
radius inclined orbits show horizon dynamics, since the on-horizon
tidal field varies as the orbit moves in the non-spherical black hole
spacetime.

The remainder of this paper is organized as follows.  We begin in
Sec.\ {\ref{sec:summary}} with a summary of the formalism that we
developed in paper I.  Section {\ref{sec:tools}} introduces the
notation and conventions that we use, and carefully defines several
quantities that are critical to our analysis, such as the
Newman-Penrose basis legs, the tidal field $\Psi_0$, and the horizon's
shear $\sigma$.  This section also briefly describes the techniques we
use to compute these quantities; further details are given in paper I.
Appendix {\ref{app:onhorizmap}} supplements this material,
demonstrating that the complex fields we use for various quantities
needed to describe the horizon's distortion are equivalent to certain
2nd-rank tensors defined on the horizon which other authors have used
(notably Ref.\ {\cite{vpm11}}, hereafter VPM11).  Section
{\ref{sec:horizongeom}} summarizes how these quantities are used to
understand the distorted horizon's geometry.

We show our results in Secs.\ {\ref{sec:tide}}, {\ref{sec:resultsI}},
{\ref{sec:resultsII}}, and {\ref{sec:resultsIII}}.  Much of the
horizon's dynamics turns out to be closely correlated to the dynamics
of the applied tidal field, so we begin in Sec.\ {\ref{sec:tide}} by
examining this tide in some detail.  We show that the vast majority of
the tide's behavior can be understood as a simple consequence of the
orbital dynamics.  There are, however, subtle features related to a
position-dependent phase and a mode-dependent amplitude correction
that must be explained with some care.  We turn to the horizon's
response proper in Sec.\ {\ref{sec:resultsI}}, carefully examining the
Schwarzschild limit, $a = 0$.  This limit is spherically symmetric, so
the horizon distortions must exhibit certain symmetries as an orbit is
inclined from equatorial to some arbitrary inclination $\theta_{\rm
  inc}$.  We demonstrate that this is the case.  This is not a
surprise, since the black hole perturbation theory code on which our
analysis is based has previously been shown to handle this limit
correctly {\cite{h00,dh06}}.  It is reassuring to see that the
modifications we made to analyze distorted horizons have not broken
this behavior.

In Sec.\ {\ref{sec:resultsII}}, we next compare certain important
aspects of the applied tidal field $\Psi_0$ to the horizon shear
$\sigma$ that arises from this field.  We first
(Sec.\ {\ref{sec:tideshearphase}}) look at the relative phase of the
tide and the shear, an analysis quite similar to one that we undertook
in paper I.  We focus for simplicity on equatorial orbits.  In the
Schwarzschild limit, the tide and the shear are very similar.  Much of
the difference between the two quantities is due to a simple temporal
offset of $\kappa^{-1} = 4M$ (where $\kappa$ is the event horizon's
surface gravity).  This offset can be understood by examining the
equation relating the ride to the shear in the frequency domain.  The
difference becomes much less simple as the black hole's spin is
increased.

It's worth noting that some of the physics associated with the offset
between the orbiting body and the horizon's distortion that we
discussed above is reproduced in the tide-shear analysis.  In
particular, we find that the shear response leads the applied tidal
field for $a = 0$, but lags it for large black hole spin --- just as
the horizon bulge always leads the orbit in Schwarzschild, but lags
the orbit for rapidly spinning Kerr.  Because the tide and the shear
are evaluated at the same coordinate radius, many ambiguities
associated with comparing the position of the horizon's bulge with the
position of the orbit disappear.  This helps to put notions of which
quantities ``lead'' and ``lag'' on a firm footing.

In the course of this analysis, we have found an interesting
oscillatory feature in the horizon's response which is most apparent
for strong-field orbits of rapidly rotating black holes.  Examining
the response of a black hole with spin $a = 0.9999M$ to the tidal
field of a strong-field eccentric orbit, we see a very strong response
near periapse with properties that closely correlate to the
near-periapsis orbital dynamics.  This is followed by about seven
cycles of low-amplitude, high-frequency oscillations in the horizon's
response.  We do not see corresponding oscillations in the tide.

Although we can estimate the frequency of these oscillations fairly
well, we have not been able to connect them to any of the frequencies
that describe this orbit or this black hole.  The rate at which the
oscillations decay also does not appear to correlate with any
timescale that we can imagine would lead to such behavior.  Having not
succeeded in coming up with a compelling explanation for this
phenomenon, for now we simply present it as an empirical finding of
our analysis, hoping that future work may offer some physical
understanding.

We conclude by examining the dynamics of horizon embeddings in
Sec.\ {\ref{sec:resultsIII}}.  For several representative cases, we
show a sequence of still images taken from an animation that combines
the behavior of the small body's orbit with the dynamics of the
horizon embedding.  Those animations can be found at the URL listed in
Ref.\ {\cite{animations}}.  Although we have endeavored to describe
the dynamics as clearly as possible using these stills, some of these
results are particularly clear when examined with the animations.  We
first consider orbits that are circular but inclined in
Sec.\ {\ref{sec:inclcirc}}, examining in detail orbits of a
Schwarzschild black hole and of a Kerr hole with $a = 0.6M$.  The
Schwarzschild results confirm our expectations from
Sec.\ {\ref{sec:resultsI}} about how the horizon should behave in this
spherically symmetric example.  The non-spherical Kerr results show
more interesting shape dynamics.  We then consider eccentric orbits in
Sec.\ {\ref{sec:eccentric}}.  As expected, the horizon's distortion
varies considerably as an orbit moves from $r_{\rm max}$ to $r_{\rm
  min}$ and back.  We examine in some detail two highly eccentric ($e
= 0.7$) orbits: one that is equatorial, and one inclined at
$\theta_{\rm inc} = 30^\circ$.  The generic case combines features
that we see from the inclined circular and the eccentric equatorial
limits.

\section{Summary of formalism}
\label{sec:summary}

\subsection{Tools, notation, and conventions}
\label{sec:tools}

All of our calculations are performed in the spacetime of a Kerr black
hole with mass $M$ and spin angular momentum $J$.  Throughout this
analysis, we work in ingoing coordinates $(v,r,\theta,\psi)$ which are
well behaved on the black hole's event horizon.  In these coordinates,
the spacetime's line element is given by
\begin{eqnarray}
ds^2 &=& -\left(1 - \frac{2Mr}{\Sigma}\right)dv^2 + 2dv\,dr -
2a\sin^2\theta\,dr\,d\psi
\nonumber\\
&-& \frac{4Mar\sin^2\theta}{\Sigma}dv\,d\psi + \Sigma\,d\theta^2
\nonumber\\
&+& \frac{(r^2 + a^2)^2 - a^2\Delta\sin^2\theta}{\Sigma}d\psi^2\;,
\label{eq:Kerr_ingoing}
\end{eqnarray}
with $a = J/M$.  (Here and throughout the paper, we use units in which
$G = 1 = c$.)  Equation (\ref{eq:Kerr_ingoing}) introduces the
functions $\Delta = r^2 - 2Mr + a^2$ and $\Sigma = r^2 +
a^2\cos^2\theta$.  The event horizon is at coordinate radius $r = r_+
= M + \sqrt{M^2 - a^2}$, the larger root of $\Delta$.  Although not
needed here, for completeness we note that ingoing coordinates are
simply related to the more commonly used Boyer-Lindquist coordinates
$(t,r,\theta,\phi)$: the coordinates $r$ and $\theta$ are identical,
and the ingoing time $v$ and angle $\psi$ are related to
Boyer-Lindquist time $t$ and angle $\phi$ via
\begin{eqnarray}
dv &=& dt + \frac{r^2 + a^2}{\Delta}dr\;,
\\
d\psi &=& d\phi + \frac{a}{\Delta}dr\;.
\end{eqnarray}

The tidal field which distorts the black hole's horizon arises from a
small body on a bound Kerr geodesic; detailed discussion of these
orbits, with an emphasis on the properties relevant to this analysis,
is given in Refs.\ {\cite{schmidt,fh09}}.  Such geodesics are
parameterized by three conserved integrals: the orbital energy $E$,
related to the spacetime's timelike Killing vector; the axial angular
momentum $L_z$, related to the spacetime's axial Killing vector; and
the Carter constant $Q$, related to the Kerr spacetime's Killing
tensor.  Once $E$, $L_z$, and $Q$ have been selected, the orbit's
motion is determined up to initial conditions.  A particularly
important feature of bound Kerr orbits is that they are triperiodic
{\cite{schmidt}}.  Each orbit has a frequency $\Omega_r$ which
describes radial oscillations, a frequency $\Omega_\theta$ which
describes polar oscillations, and a frequency $\Omega_\phi$ which
describes rotations about the black hole's spin axis.  Once $E$,
$L_z$, and $Q$ are known, it is not too difficult to compute
$\Omega_r$, $\Omega_\theta$, and $\Omega_\phi$ {\cite{schmidt,fh09}}.

We remap the motion in $r$ and $\theta$ to the parameters $p$
(semi-latus rectum), $e$ (eccentricity), and $\theta_{\rm m}$, defined
by
\begin{eqnarray}
r &=& \frac{p}{1 + e\cos\chi_r}\;,
\label{eq:rofpsi}\\
\cos\theta &=& \cos\theta_{\rm m}\cos(\chi_\theta +
\chi_{\theta,0})\;.
\label{eq:thetaofchi}
\end{eqnarray}
The orbit's radius $r$ thus oscillates between periapsis $r_{\rm min}
= p/(1 + e)$ and apoapsis $r_{\rm max} = p/(1 - e)$; the polar angle
$\theta$ oscillates between $\theta_{\rm min} = \theta_{\rm m}$ and
$\theta_{\rm max} = 180^\circ - \theta_{\rm m}$.

With the parameterization (\ref{eq:rofpsi}) and (\ref{eq:thetaofchi}),
the geodesic equations for the coordinates $r$ and $\theta$ become
equations for the angles $\chi_r$ and $\chi_\theta$.  Note that we
could include an offset phase $\chi_{r,0}$ in Eq.\ (\ref{eq:rofpsi}).
We have set $\chi_{r,0} = 0$, which is equivalent to setting the
origin of our time coordinate to the moment that the orbit passes
through periapsis.  References {\cite{schmidt,fh09}} give easy-to-use
expressions relating the ($E$, $L_z$, $Q$) and ($p$, $e$, $\theta_{\rm
  m}$) parameterizations.  For much of our analysis, we use the angle
$\theta_{\rm inc}$ introduced in Ref.\ {\cite{dh06}} in place of
$\theta_{\rm m}$:
\begin{eqnarray}
\theta_{\rm inc} = 90^\circ - {\rm sgn}(L_z)\theta_{\rm m}\;.
\label{eq:thetainc}
\end{eqnarray}
This angle varies smoothly from $0$ to $180^\circ$ as the orbit varies
from prograde equatorial ($\theta_{\rm m} = 90^\circ$, $L_z > 0$) to
retrograde equatorial ($\theta_{\rm m} = 90^\circ$, $L_z < 0$).

The tidal field is quantified by the complex scalar field\footnote{In
  this paper, we use capital $\Psi$ rather than the more commonly used
  lowercase $\psi$ to denote the Weyl curvature scalars in order to
  avoid confusion with the ingoing axial coordinate.} $\Psi_0$, which
is built from the Weyl curvature tensor:
\begin{equation}
\Psi_0 = -C_{\mu\alpha\nu\beta}l^\mu m^\alpha l^\nu m^\beta\;.
\label{eq:psi0def}
\end{equation}
The vectors used here are the Newman-Penrose null legs in the
Hawking-Hartle representation {\cite{hh72}}:
\begin{eqnarray}
l^\mu &\doteq& \left[1, \frac{\Delta}{2\varpi^2}, 0,
  \frac{a}{\varpi^2}\right]\;,
\label{eq:lHH}
\\
n^\mu &\doteq& \frac{1}{\Sigma}\biggl[-a^2\sin^2\theta/2, -\varpi^2 +
  \frac{a^2\Delta\sin^2\theta}{4\varpi^2}, 0,
\nonumber\\
& &\qquad -a + \frac{a^3\sin^2\theta}{2\varpi^2}\biggr]\;,
\label{eq:nHH}
\\
m^\mu &\doteq& \frac{1}{\sqrt{2}(r + ia\cos\theta)}\biggl[0,
  -\frac{ia\Delta\sin\theta}{\varpi^2}, 1,
\nonumber\\
& &\qquad i\csc\theta - \frac{ia^2\sin\theta}{\varpi^2}\biggr]\;.
\label{eq:mHH}
\end{eqnarray}
The symbol $\doteq$ means ``the components of the quantity on the
left-hand side are represented by the array on the right-hand side in
ingoing Kerr coordinates.''  For brevity, we have introduced $\varpi^2
= r^2 + a^2$.  These legs satisfy
\begin{equation}
l^\mu n_\mu = -1\;,\qquad m^\mu \bar m_\mu = 1\;,
\end{equation}
with overbar denoting complex conjugate; all other inner products
between legs vanish.

Following VPM11, the Weyl curvature on the horizon is completely
described by a two-dimensional trace-free symmetric tensor $C_{AB}$,
where capital Roman indices denote components associated with
coordinates on the horizon.  Such a tensor has only two independent
components, which we can describe as ``tidal polarizations,'' and
denote $C_+$ and $C_\times$.  These polarizations are simply related
to the curvature scalar $\Psi_0$ on the horizon:
\begin{equation}
\Psi_0(r_+) = -\left(C_+ + iC_\times\right)\;.
\label{eq:tidalpolarizations}
\end{equation}
See App.\ {\ref{app:onhorizmap}} for further details and a proof of
Eq.\ (\ref{eq:tidalpolarizations}).  We use the polarizations
$C_{+,\times}$ in much of our presentation of results, especially in
Sec.\ {\ref{sec:resultsII}}.

The tidal field can be decomposed into harmonics of the three
fundamental Kerr frequencies, allowing us to write its value at $r =
r_+$ as
\begin{equation}
\Psi_0(v,\theta,\psi) = \frac{1}{16M^2r_+^2}\sum_{lmkn}
W^{\rm H}_{lmkn}S^+_{lmkn}(\theta) e^{i\Phi_{mkn}(v,\psi)}\;.
\label{eq:Psi0expand}
\end{equation}
The function
\begin{equation}
S^+_{lmkn}(\theta) = {_{+2}}S_{lm}(\theta;a\omega_{mkn})
\end{equation}
is a spheroidal harmonic of spin-weight $+2$; detailed discussion of
this function and how it is computed can be found in
Ref.\ {\cite{h00}}.  The frequency $\omega_{mkn}$ is a harmonic of the
orbital frequencies,
\begin{equation}
\omega_{mkn} = m\Omega_\phi + k\Omega_\theta + n\Omega_r\;.
\end{equation}
The product $a\omega_{mkn}$ sets the ``oblateness'' associated with
$S^+_{lmkn}(\theta)$.  We describe the phase $\Phi_{mkn}(v,\psi)$ in
more detail below.

The amplitude $W^{\rm H}_{lmkn}$ can be found by solving the Teukolsky
equation {\cite{t73}}.  In practice, we compute the field $\Psi_4$, a
different projection of the Weyl curvature.  In the limits $r \to r_+$
and $r \to \infty$, the fields $\Psi_4$ and $\Psi_0$ can be related to
one another without too much trouble {\cite{tp74}}.  As $r \to r_+$,
$\Psi_4$ takes the form
\begin{equation}
\Psi_4 = \frac{\Delta^2}{\left(r - ia\cos\theta\right)^4}
\sum_{lmkn}Z^{\rm H}_{lmkn}S^-_{lmkn}(\theta)e^{i\Phi_{mkn}(v,\psi)}\;.
\label{eq:Psi4expand}
\end{equation}
Detailed discussion of how to compute the amplitude $Z^{\rm H}_{lmkn}$
using the Teukolsky equation is given in Ref.\ {\cite{dh06}}.  The
function
\begin{equation}
S^-_{lmkn}(\theta) = {_{-2}}S_{lm}(\theta;a\omega_{mkn})
\end{equation}
is a spheroidal harmonic of spin-weight $-2$; see {\cite{h00}} for
detailed discussion.

The Starobinsky-Churilov identities {\cite{sc73}} connect the
amplitudes of these two curvature scalars:
\begin{equation}
W^{\rm H}_{lmkn} = \beta_{lmkn}Z^{\rm H}_{lmkn}\;,
\label{eq:sc_identity}
\end{equation}
where
\begin{eqnarray}
\beta_{lmkn} &=& \! \frac{64(2Mr_+)^4p_{mkn}(p_{mkn}^2 + \kappa^2)(p_{mkn}
  + 2i\kappa)}{c_{lmkn}}\;,
\nonumber\\
\label{eq:betalmkndef}\\
|c_{lmkn}|^2 &=& \{[(\lambda + 2)^2 + 4ma\omega_{mkn} - 4a^2\omega_{mkn}^2]
\nonumber\\
& &\times(\lambda^2 + 36ma\omega_{mkn} - 36a^2\omega_{mkn}^2)
\nonumber\\
& &+(2\lambda + 3)(96a^2\omega_{mkn}^2 - 48ma\omega_{mkn})\}
\nonumber\\
& &+144\omega_{mkn}^2(M^2 - a^2)\;,
\\
{\rm Im}\,c_{lmkn} &=& (-1)^{l+k+m}\,12M\omega_{mkn}\;,
\\
{\rm Re}\,c_{lmkn} &=& +\sqrt{|c_{lmkn}|^2 - 144M^2\omega_{mkn}^2}\;.
\end{eqnarray}
In these equations,
\begin{equation}
p_{mkn} = \omega_{mkn} - m\Omega_{\rm H}\;,
\end{equation}
with $\Omega_{\rm H} = a/2Mr_+$, the angular frequency associated with
the Kerr event horizon.  The real number $\lambda$ is related to the
eigenvalue of the spheroidal harmonic:
\begin{equation}
\lambda = {\cal E} - 2am\omega_{mkn} + a^2\omega_{mkn}^2 - s(s+1)\;,
\end{equation}
where $s = -2$, and ${\cal E}$ is the eigenvalue\footnote{Multiple
  conventions for this eigenvalue can be found in the literature.
  Another common one puts $\lambda = A - 2am\omega_{mkn} +
  a^2\omega_{mkn}^2$; they are related by $A = {\cal E} - s(s+1)$.}
associated with the $s = -2$ spheroidal harmonic.  In the limit $a =
0$, ${\cal E} = l(l+1)$.  Note that the imaginary part of $c_{lmkn}$
is positive for ``polar'' modes ($l + k + m$ even), and is negative
for ``axial'' modes ($l + k + m$ odd).  This sign is given incorrectly
in many papers in the literature, including the first one in which the
constant is computed {\cite{tp74}}.  We discuss this error briefly in
an erratum which was recently published for Ref.\ {\cite{paperI}};
further discussion will be given in a forthcoming paper by Flanagan
and Hinderer {\cite{flanhind_inprep}}.

Other important quantities appearing in these equations are the black
hole's surface gravity,
\begin{equation}
\kappa = \frac{\sqrt{M^2 - a^2}}{2Mr_+}\;,
\end{equation}
and the phase
\begin{equation}
\Phi_{mkn}(v,\psi) = m[\psi - K(a)] - (m\Omega_\phi + k\Omega_\theta +
n\Omega_r)v\;,
\label{eq:Phimkn}
\end{equation}
where
\begin{eqnarray}
K(a) &=& \frac{a}{2M(Mr_+ - a^2)}\biggl\{a^2 - Mr_+
\nonumber\\
&+& 2M^2{\rm arctanh}\left(\sqrt{1 - a^2/M^2}\right)
\nonumber\\
&+& M\sqrt{M^2 - a^2} \ln\left[\frac{a^2}{4(M^2 - a^2)}\right]\biggr\}\;.
\label{eq:K_of_a}
\end{eqnarray}

In Eqs.\ (\ref{eq:Psi0expand}) and (\ref{eq:Psi4expand}), the sum over
$l$ goes from $2$ to $\infty$, the sum over $m$ from $-l$ to $l$, and
the sums over $k$ and $n$ from $-\infty$ to $\infty$.  We abbreviate
this set of indices $\Lambda \equiv \{l,m,k,n\}$.  With this,
Eq.\ (\ref{eq:Psi0expand}) becomes
\begin{eqnarray}
\Psi_0(v,\theta,\psi) &=& \frac{1}{16M^2 r_+^2} \sum_{\Lambda}
W^{\rm H}_{\Lambda}S^+_{\Lambda}(\theta) e^{i\Phi_{\Lambda}(v,\psi)}
\nonumber\\
&\equiv& \sum_{\Lambda} \Psi_{0,\Lambda} S^+_{\Lambda}(\theta)
e^{i\Phi_{\Lambda}(v,\psi)}\;.
\label{eq:Psi0expand2}
\end{eqnarray}
We have introduced
\begin{eqnarray}
\Psi_{0,\Lambda} &=& \frac{W^{\rm H}_\Lambda}{16M^2r_+^2}
\nonumber\\
&=&
\frac{64M^2r_+^2p_\Lambda(p^2_\Lambda + \kappa^2)(p_\Lambda +
  2i\kappa)Z^{\rm H}_\Lambda}{c_\Lambda}\;.
\label{eq:Psi0mode}
\end{eqnarray}
Note that the phase $\Phi_{\Lambda} \equiv \Phi_{mkn}$ and wavenumber
$p_\Lambda \equiv p_{mkn}$ don't actually depend on the index $l$.
Using $\Lambda$ as a label for these quantities is thus somewhat
redundant, though this redundancy is harmless.

\subsection{The geometry of a distorted event horizon}
\label{sec:horizongeom}

\subsubsection{The shear to the horizon's generators}

The first tool we need to understand how the tidal field affects the
horizon's geometry is the shear $\sigma$ of the horizon's generators.
It is given by
\begin{equation}
\sigma = m^\mu m^\nu \nabla_\mu l_\nu\;,
\label{eq:sheardef}
\end{equation}
evaluated at $r = r_+$.  (Note that, for an unperturbed black hole,
$l^\mu$ is tangent to the generators at $r = r_+$.)  Just as the
complex Weyl scalar $\Psi_0$ can be written using polarizations
$C_{+,\times}$ of the on-horizon Weyl tensor, the complex shear can be
written in terms of polarizations $\sigma_{+,\times}$ of an on-horizon
shear tensor:
\begin{equation}
\sigma = \sigma_+ + i\sigma_\times\;.
\label{eq:shearpolarizations}
\end{equation}
See App.\ {\ref{app:onhorizmap}} for further details and a proof of
Eq.\ (\ref{eq:shearpolarizations}).  We will use $\sigma_{+,\times}$
in much of our discussion of results, especially in
Sec.\ {\ref{sec:resultsII}}.

With the tetrad and gauge that we use, the perturbed shear is governed
by the equation {\cite{h74}}
\begin{equation}
(D - \kappa)\sigma = \Psi_0\;,
\label{eq:shearevolve}
\end{equation}
where the derivative operator $D \equiv l^\mu\partial_\mu$.  Let us
expand $\sigma$ as we expanded $\Psi_0$:
\begin{equation}
\sigma(v,\theta,\psi) = \sum_{\Lambda}
\sigma_{\Lambda}S^+_{\Lambda}(\theta) e^{i\Phi_{\Lambda}(v,\psi)}\;.
\label{eq:shearexpand}
\end{equation}
Using the fact that $D \to \partial_v + \Omega_{\rm H}\partial_\psi$
on the horizon, we find that Eq.\ (\ref{eq:shearevolve}) is satisfied
if
\begin{equation}
\sigma_{\Lambda} = 64M^2r_+^2c_\Lambda^{-1}p_\Lambda(p_\Lambda + i\kappa)
(ip_\Lambda - 2\kappa) Z^{\rm H}_\Lambda\;.
\nonumber\\
\label{eq:shearamplitude}
\end{equation}
As was extensively discussed in paper I, there is a phase offset
between the shear and the applied tidal field.  The phase offset for
each mode is simple to calculate:
\begin{equation}
\frac{\sigma_\Lambda}{\Psi_{0,\Lambda}} = \frac{i}{p_\Lambda -
  i\kappa} =
\frac{\exp\left[-i\arctan(p_\Lambda/\kappa)\right]}{\sqrt{p_\Lambda^2
    + \kappa^2}}\;.
\label{eq:modalphaseoffset}
\end{equation}
In other words, for each mode, the shear leads the tide by an angle
given by the mode's wavenumber $p_\Lambda$ times the inverse surface
gravity $\kappa^{-1}$.  For circular and equatorial orbits, $p_\Lambda
\to m(\Omega_\phi - \Omega_{\rm H})$, so each mode experiences the
same phase shift, modulo $m$.  For these orbits, we find a simple,
constant offset between the tidal field and the resulting shear.  More
complicated behavior results for generic orbits, since many modes,
each with different phase shifts, contribute to $\Psi_0$ and $\sigma$.

Although we do all of our calculations in this paper in the frequency
domain, it is also useful to examine Eq.\ (\ref{eq:shearevolve}) in a
``time-like'' domain.  As mentioned above, the Newman-Penrose leg
$l^\mu$ is tangent to the unperturbed horizon generators at $r = r_+$.
We may therefore write $D \equiv d/d\lambda$ on the horizon, where
$\lambda$ is affine parameter along the generator.  In this
representation, $\lambda$ is effectively a time measure, albeit a
somewhat unusual time, measured by a clock that ticks at a uniform
rate as it follows a specific horizon generator.

With this in mind, following Ref.\ {\cite{membrane}} Sec.\ VI C 6, let
us find the Green's function $G(\lambda,\lambda')$ for
Eq.\ (\ref{eq:shearevolve}):
\begin{equation}
(D - \kappa)G(\lambda,\lambda') = \delta(\lambda - \lambda')\;.
\end{equation}
This equation has the solution
\begin{equation}
G(\lambda,\lambda') = -e^{\kappa(\lambda - \lambda')}\Theta(\lambda' -
\lambda)\;,
\end{equation}
where the step function
\begin{eqnarray}
\Theta(x) &=& 1 \qquad x > 0
\nonumber\\
&=& 0 \qquad x < 0\;.
\end{eqnarray}
The shear along the generator is then
\begin{equation}
\sigma(\lambda) = -\int_\lambda^\infty e^{\kappa(\lambda -
  \lambda')}\Psi_0(\lambda')d\lambda'\;,
\end{equation}
or, using Eqs.\ (\ref{eq:tidalpolarizations}) and
(\ref{eq:shearpolarizations}),
\begin{equation}
\sigma_{+,\times}(\lambda) = \int_\lambda^\infty e^{\kappa(\lambda -
  \lambda')}C_{+,\times}(\lambda')d\lambda'\;.
\label{eq:sheargreens}
\end{equation}
Notice that the behavior at $\lambda$ depends on the tides {\it to the
  future} of $\lambda$ --- a manifestation of the horizon's
teleological nature.  What we see is that the shear at $\lambda$ on a
particular generator depends on the tide integrated over an interval
from $\lambda$ to $\lambda + \mbox{(a few)}\times\kappa^{-1}$.

\subsubsection{The curvature of the distorted horizon}

The tidal field $\Psi_0$ on the horizon also tells us the scalar Ricci
curvature of the black hole, $R_{\rm H}$.  This is discussed in great
detail in paper I.  Briefly, the scalar curvature of the hole's event
horizon is given by
\begin{equation}
R_{\rm H} = R^{(0)}_{\rm H} + R^{(1)}_{\rm H}\;,
\end{equation}
where
\begin{equation}
R^{(0)}_{\rm H} = \frac{2}{r_+^2}\frac{(1 + a^2/r_+^2)(1
  -3a^2\cos^2\theta/r_+^2)}{(1 + a^2\cos^2\theta/r_+^2)^3}
\end{equation}
describes an undistorted Kerr black hole, and
\begin{equation}
R^{(1)}_{\rm H} = -4{\rm Im}\sum_\Lambda \frac{\bar\eth\bar\eth
  \Psi_{0,\Lambda}}{p_\Lambda(ip_\Lambda + \kappa)}
\end{equation}
is the perturbation to $R_{\rm H}$ arising from the tidal field
$\Psi_0$.  The operator $\bar\eth$ lowers the spin weight of the
angular basis functions.  As discussed in Sec.\ IIC of paper I, it is
quite simple to evaluate $\bar\eth\bar\eth\Psi_{0,\Lambda}$ with the
spectral expansion for the spin-weighted spheroidal harmonics that we
use.  See paper I for detailed discussion.

To visualize the curvature of a distorted horizon, we embed the
horizon in a global Euclidean 3-space.  This means finding the
function
\begin{equation}
r_{\rm E}(\theta,\psi) = r_{\rm E}^{(0)}(\theta) + r_{\rm
  E}^{(1)}(\theta,\psi)
\end{equation}
that defines a surface with the same Ricci scalar curvature as the
distorted horizon.  This works well for spins $a/M < \sqrt{3}/2$; for
higher spins, global Euclidean embeddings do not exist even for the
undistorted event horizon {\cite{smarr}}.  As such, we confine our
embedding visualizations in this paper to the range $0 \le a/M <
\sqrt{3}/2$.  Work in progress indicates that an elegant way to lift
this restriction will to be embed the horizon's distorted geometry in
the globally hyperbolic space $H^3$ {\cite{ghr09}}.

Confining our discussion to Euclidean 3-space, a simple analytic
solution exists for the undistorted hole's embedding radius $r_{\rm
  E}^{(0)}(\theta)$ {\cite{smarr}}.  To find the perturbation
$r^{(1)}_{\rm E}(\theta,\psi)$, we expand in spherical harmonics,
writing
\begin{equation}
r^{(1)}_{\rm E}(\theta,\psi) = r_+\sum_{\ell m}\varepsilon_{\ell m}
Y_{\ell m}(\theta, \psi)\;.
\end{equation}
Given this form, it is a straightforward (although rather lengthy)
exercise to construct the scalar curvature $R^{(1)}_{\rm E}$
associated with $r^{(1)}_{\rm E}$; details are given in Appendix B of
paper I.  By enforcing $R^{(1)}_{\rm E} = R^{(1)}_{\rm H}$, we read
off the embedding coefficients $\varepsilon_{\ell m}$.  Full details
of the algorithm for doing this are given in Appendix B of paper I.

\section{Behavior of the tidal field}
\label{sec:tide}

Before examining how the horizon responds to dynamical tides, we first
look at some examples of tides from geodesic orbits.  As we will see
in later sections, the shear polarizations $\sigma_{+,\times}$ largely
follow the behavior of the driving tides $C_{+,\times}$.  There are,
however, some features of the shear that are unique.  It is thus
useful to examine the tide in detail to set a baseline for comparing
the two functions' behaviors.

\subsection{Circular inclined orbits}
\label{sec:circtide}

\begin{figure}[ht]
\includegraphics[width = 0.48\textwidth]{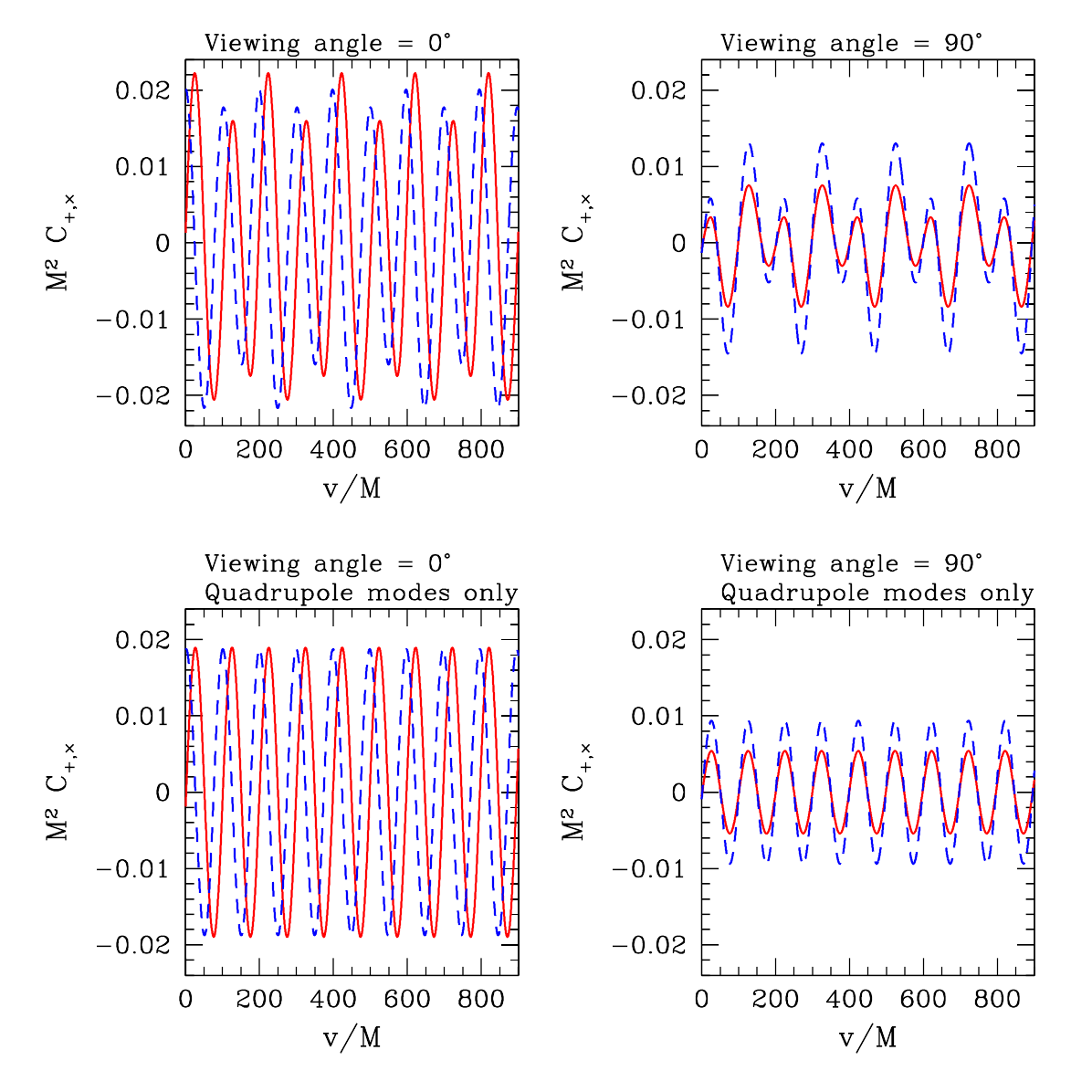}
\caption{Examples of the on-horizon tidal field $C$ for circular
  orbits around a Schwarzschild black hole.  The four plots shown here
  are for an orbit with $p = 10M$ and with inclination angle
  $\theta_{\rm inc} = 60^\circ$.  The solid (red) line shows the
  polarization $C_+$; the dashed (blue) line shows $C_\times$.  The
  top two panels illustrate how this field varies with ingoing time
  $v$ at azimuth angle $\psi = 0^\circ$ and $\psi = 90^\circ$; the
  bottom two figures show the same data, but including only
  quadrupolar modes (i.e., modes for which $l = 2$ and $m + k = \pm
  2$).  These plots illustrate the importance of modes beyond the
  quadrupole, as well as the strong functional dependence of the tide
  on the position at which it is measured.}
\label{fig:schwpsi0_circ}
\end{figure}

The simplest behavior is seen for circular orbits of Schwarzschild
black holes, for which $\Omega_\theta = \Omega_\phi$.  These orbits do
not precess, instead maintaining a fixed orientation for all time.
Figure {\ref{fig:schwpsi0_circ}} shows\footnote{We use a mass ratio
  $\mu/M = 1/30$ for all figures which show quantities computed in
  black hole perturbation theory (such as $C_+$, $\sigma_+$, or the
  embedding surfaces).  This fairly large mass ratio is only used so
  that the effects we compute are clearly visible in these figures.
  Since we use linear perturbation theory, one can easily extrapolate
  to other mass ratios.  We also scale these quantities by $r_{\rm
    min}^3/M^3 = p^3/[M(1 + e)]^3$, accounting for the $1/r^3$
  leading-order scaling associated with tides.  This makes it easier
  to compare tidal distortions for different orbits, ensuring that the
  maximum distortion is roughly the same in all cases we present.} a
typical example of the behavior we see in this case.  The four panels
of this figure all illustrate the tidal field arising from an orbit
with $a = 0$, $p = 10M$, $\theta_{\rm inc} = 60^\circ$.  The orbit is
oriented so that $\theta = 30^\circ$ at $\psi = 0^\circ$.  It crosses
the equator at $\psi = 90^\circ$, continues to $\theta = 150^\circ$ at
$\psi = 180^\circ$, crosses the equator again at $\psi = 270^\circ$
and returns to $\theta = 30^\circ$ at $\psi = 360^\circ$.

Although simple, the tidal field shown in
Fig.\ {\ref{fig:schwpsi0_circ}} demonstrates certain important
features which will recur in more complicated examples.  Perhaps most
significantly, note the strong modulation of the tide's amplitude with
azimuthal position $\psi$.  The panels on the left show the tide
evaluated at $\psi = 0^\circ$, where the tide is near its maximum;
those on the right show it at $\psi = 90^\circ$, near its minimum.
The amplitude varies sinusoidally with $\psi$ between these extremes.

The two upper panels show the tidal field including modes up to $l =
9$; the two lower panels only include quadrupole modes ($l = 2$, $m +
k = \pm 2$).  The quadrupolar tidal field is a pure sinusoid,
oscillating twice per orbit.  Additional modes complicate this
structure, adding features which oscillate at both lower frequency ($m
= 1$ modes) and higher frequency.

\begin{figure*}[ht]
\includegraphics[width = 0.48\textwidth]{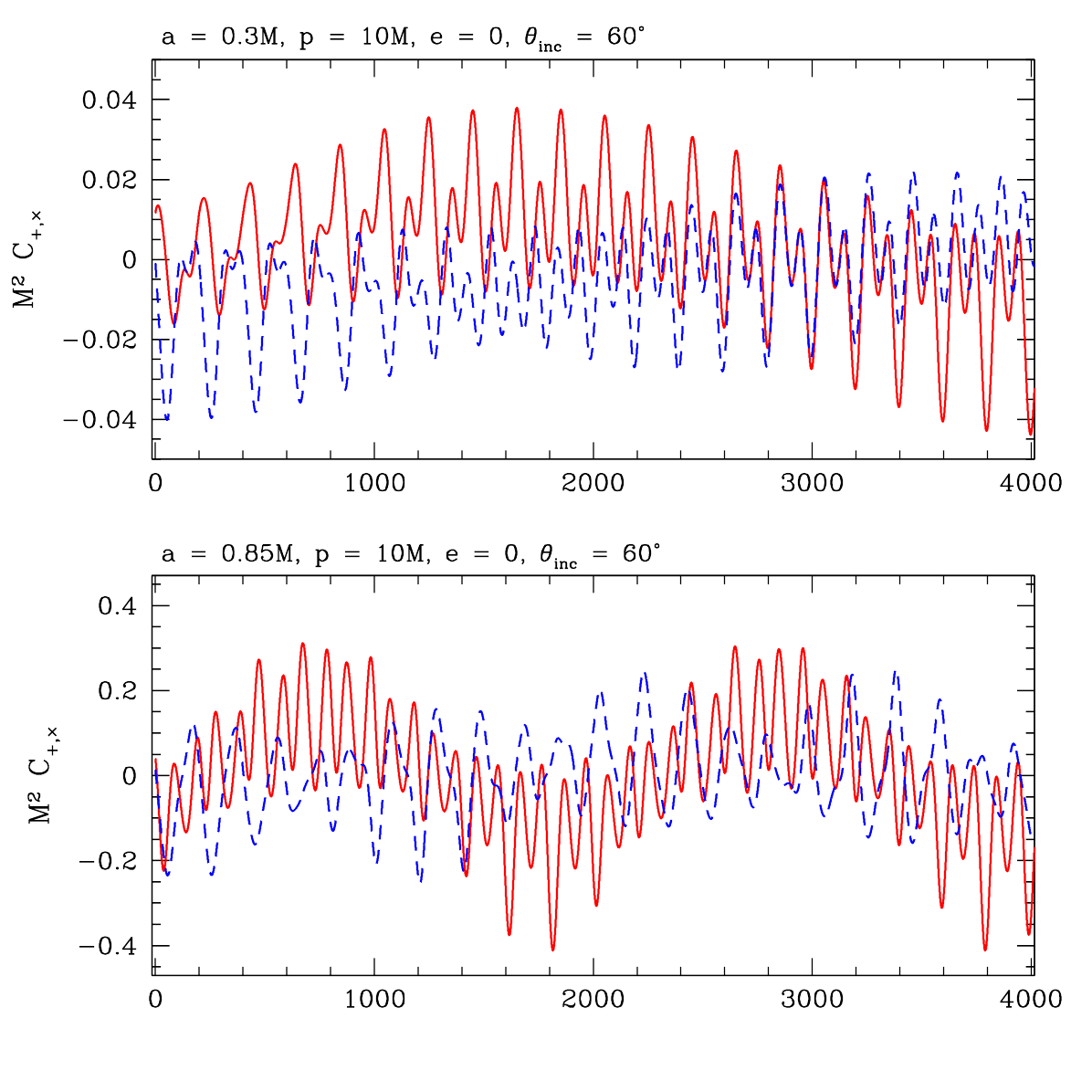}
\includegraphics[width = 0.48\textwidth]{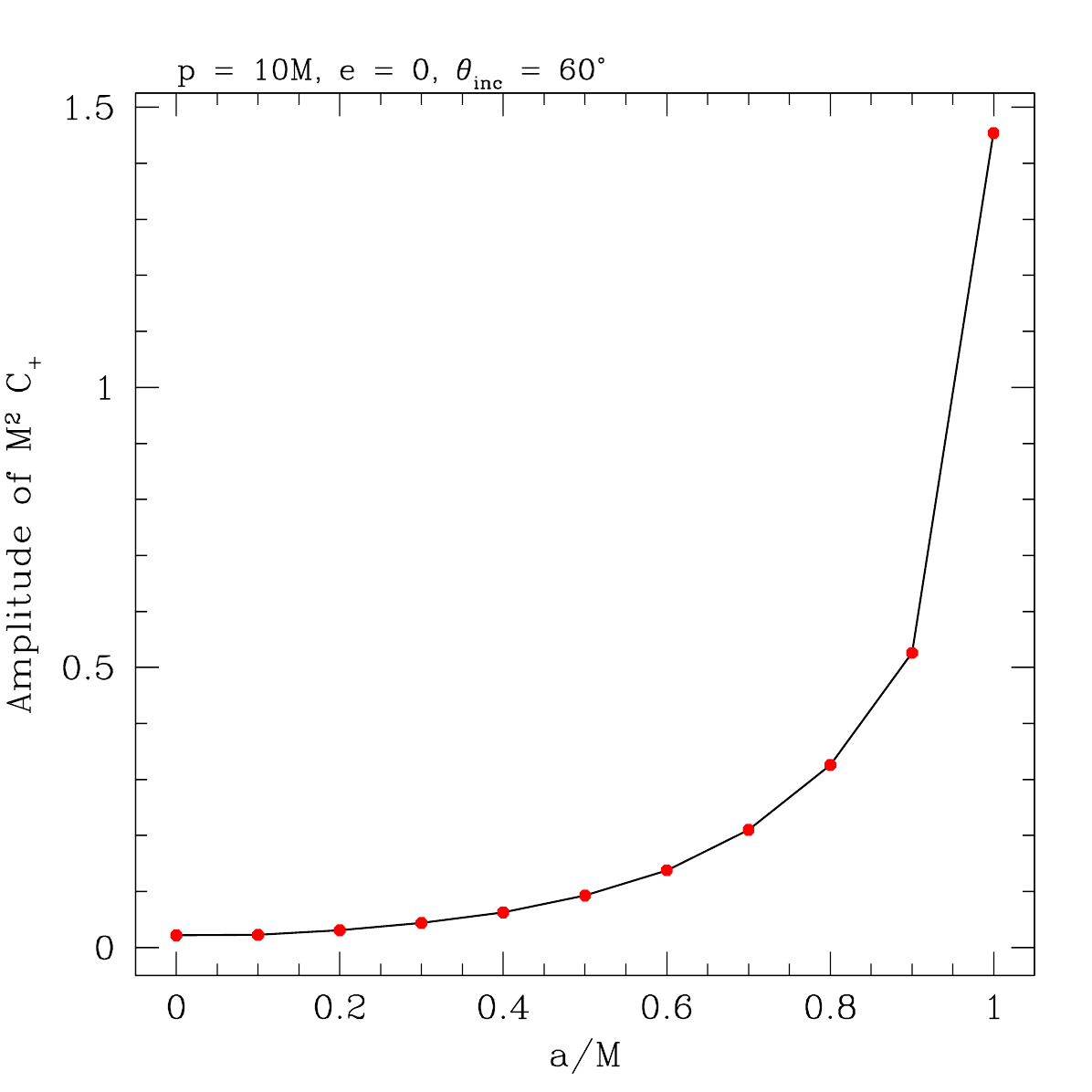}
\caption{Examples of the on-horizon tidal polarizations $C_{+,\times}$
  for circular orbits around Kerr black holes (left panels), and the
  amplitude as a function of black hole spin (right panel).  All data
  are for orbits with $p = 10M$, $\theta_{\rm inc} = 60^\circ$.  The
  top plot on the left is for a black hole with spin $a = 0.3M$;
  bottom is for a hole with spin $a = 0.85M$.  Notice that the
  amplitude is significantly smaller for $a = 0.3M$, and that the
  modulation is significantly slower.  The difference in modulation is
  simply explained: frame dragging is substantially stronger at $a =
  0.85M$, so the orbit precesses much more rapidly.  The amplitude
  effect is more subtle.  Each mode $\Lambda$ of the on-horizon tidal
  field is proportional to $p_\Lambda \equiv \omega_\Lambda -
  m\Omega_{\rm H}$.  At $a = 0.3M$, both $\Omega_\phi$ and
  $\Omega_\theta$ are close enough to $\Omega_{\rm H}$ to suppress the
  most important modes of the tide.  On the right, we we show the
  amplitude of $C_+$ as a function of $a$, showing how strongly this
  field varies thanks to this proportionality of the modes with
  $p_\Lambda$.}
\label{fig:kerrpsi0_circ}
\end{figure*}

Consider next the tide arising from circular orbits of Kerr black
holes.  Two examples are shown in the left-hand panels of
Fig.\ {\ref{fig:kerrpsi0_circ}}.  The top example is for spin $a =
0.3M$, and the bottom is for $a = 0.85M$; both examples use $p = 10M$,
$\theta_{\rm inc} = 60^\circ$ and include modes up to $l = 9$.  Thanks
to frame dragging, the orbit's orientation is not fixed in these
cases.  The orbit instead precesses about the black hole's spin axis,
modulating the tide.  This precession causes a modulation of the
fields $C_{+,\times}$; they oscillate between bounds similar to those
seen in Fig.\ {\ref{fig:schwpsi0_circ}} at $\psi = 0^\circ$ and $\psi
= 90^\circ$.  This precession is substantially faster at $a = 0.85M$
than at $a = 0.3M$, leading to the more rapid modulation seen in the
bottom figure than in the top.

More interestingly, the amplitude is roughly an order of magnitude
smaller for $a = 0.3M$ than for $a = 0.85M$.  The reason for this can
be understood by examining Eq.\ (\ref{eq:Psi0mode}): each mode of the
tidal field is proportional to the wavenumber $p_\Lambda =
\omega_\Lambda - m\Omega_{\rm H}$.  For $a = 0.3M$, $\Omega_\theta$
and $\Omega_\phi$ are roughly a factor of two from $\Omega_{\rm H}$
($M\Omega_\theta = 0.0312$, $M\Omega_\phi = 0.0317$, $M\Omega_{\rm H}
= 0.0768$).  By contrast, these frequencies are quite different for $a
= 0.85M$ ($M\Omega_\theta = 0.0303$, $M\Omega_\phi = 0.0318$,
$M\Omega_{\rm H} = 0.2784$).  The wavenumber is substantially smaller
in the case $a = 0.3M$ for the most important modes of the tide, and
the resulting field is of much smaller amplitude.  This dependence of
tide on $p_\Lambda$ causes a strong variation of its amplitude as a
function of $a$.  The right-hand panel of
Fig.\ {\ref{fig:kerrpsi0_circ}} shows how, holding the orbit geometry
fixed, the amplitude of $C_+$ varies with $a$.  The effect is quite
significant, with the field being a factor $\sim 70$ larger for nearly
maximal Kerr holes than it is for very slow rotation.  Although
differing in detailed behavior, similar variation of the tide with $a$
is found for other orbits.  For example, for strong-field orbits with
($p, e, \theta_{\rm inc}$) fixed, the we typically find a minimal tide
at $a/M \sim 0.1 - 0.3$.

\subsection{Eccentric equatorial orbits}

\begin{figure*}[ht]
\includegraphics[width = 0.48\textwidth]{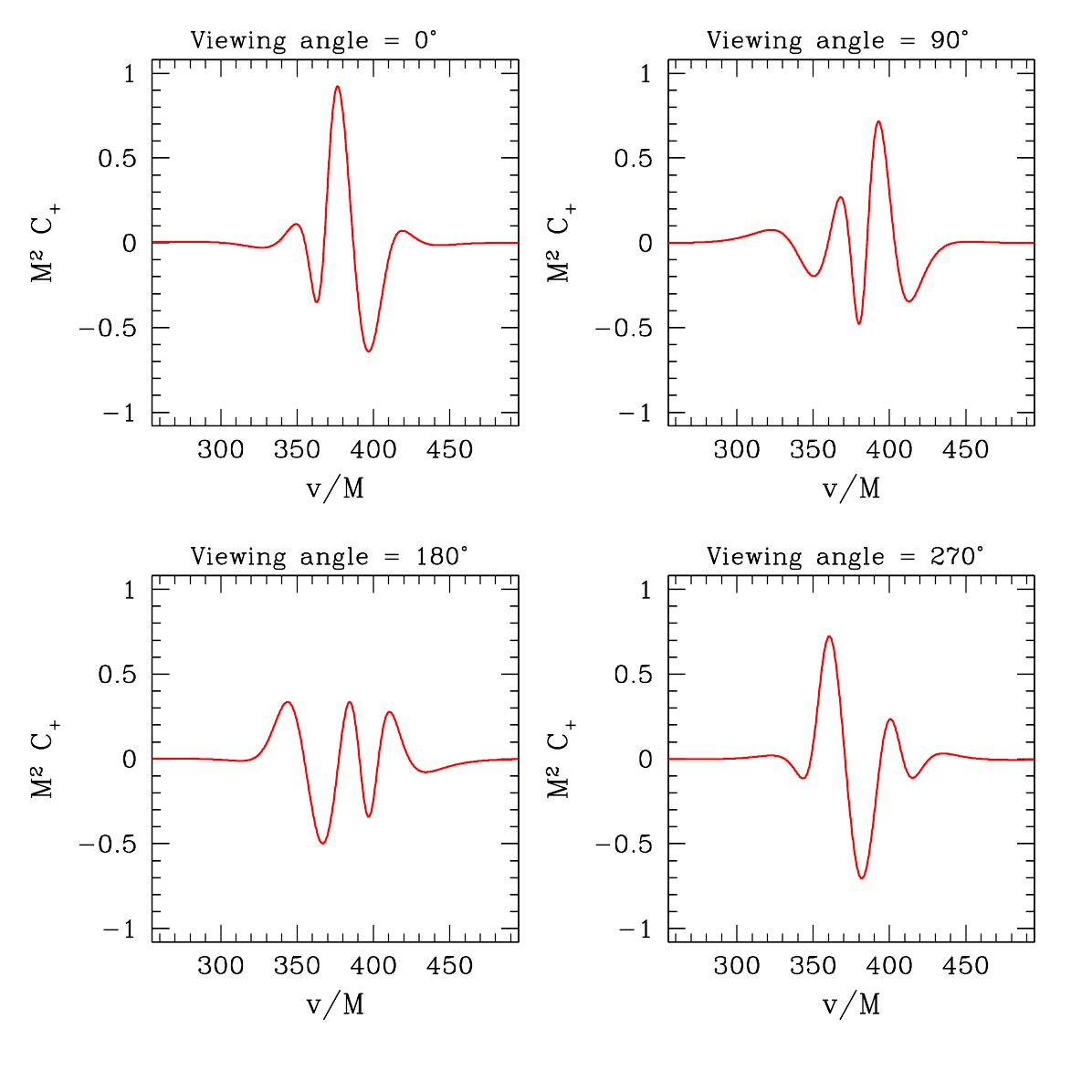}
\includegraphics[width = 0.48\textwidth]{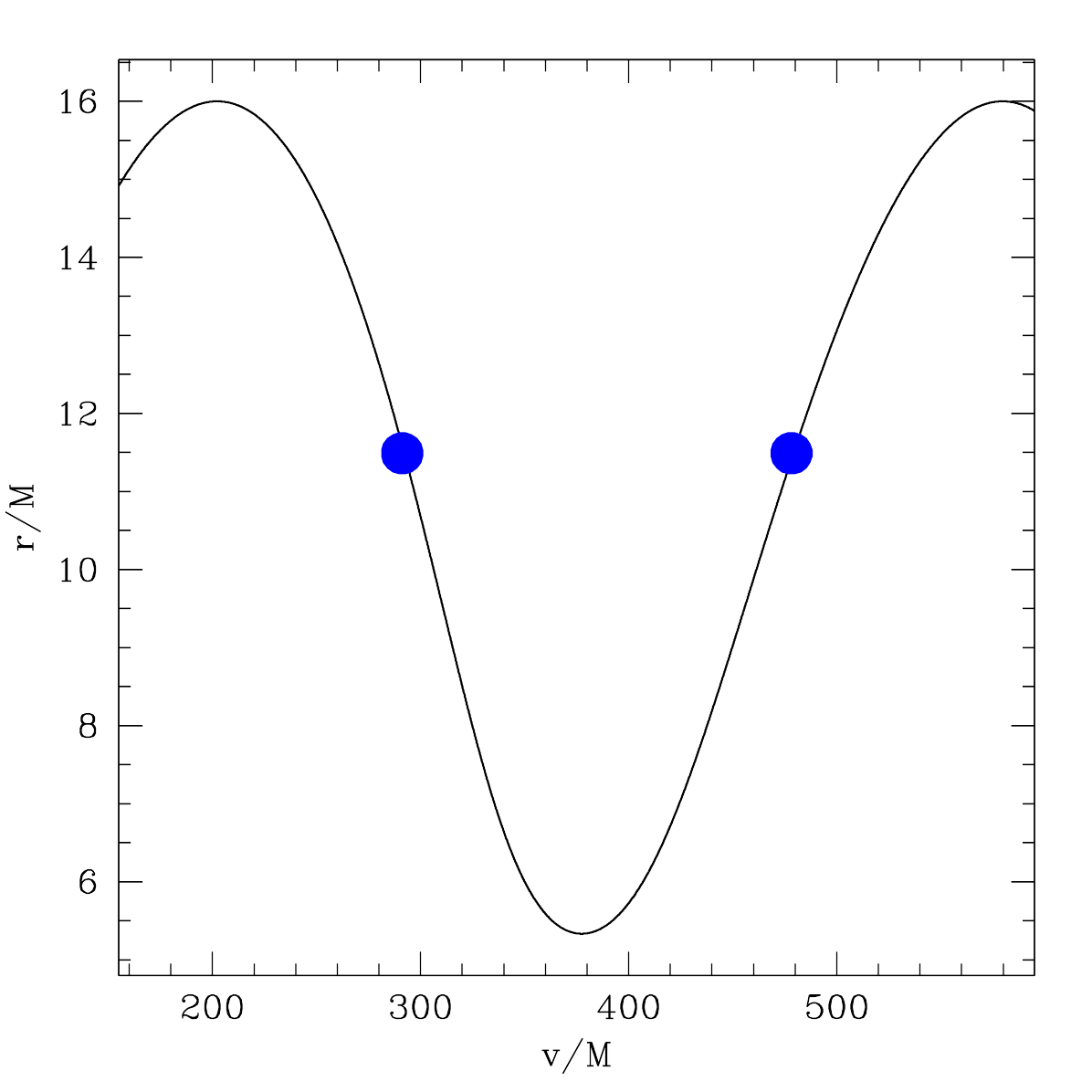}
\caption{Examples of the on-horizon tidal field $C_+$ for an eccentric
  equatorial orbit of a Schwarzschild black hole (left), and a portion
  of that orbit's radial motion (right). The orbit has $p = 8M$, $e =
  0.5$.  We examine $C_+$ on the hole's equator at four different
  axial positions.  (We don't show the cross polarization since
  $C_\times = 0$ at $\theta = 90^\circ$ for tides from equatorial
  orbits.)  Consider first the tidal field at $\psi = 0^\circ$.  The
  shape of this field can be considered a prototype of the tide from
  an eccentric orbit: There is a large spike at roughly the moment the
  orbit passes through $r = r_{\rm min}$, with smaller scale
  oscillations before and after.  The oscillations and spike occur
  only when the orbit is close to the event horizon.  In the
  right-hand panel, we have marked with large dots the moments when
  the orbital radius is $r = 10^{1/3}r_{\rm min}$.  Since the tide
  scales roughly as $1/r^3$, these should indicate when the tide has
  fallen by about an order of magnitude from its peak value.  Indeed,
  at these moments ($v \simeq 295M$ and $v\simeq 475M$) the tide has
  fallen by at least an order of magnitude.  The three other small
  panels on the left-hand side illustrate how the tide is modulated by
  azimuth angle.  The position at which the field is measured can
  significantly affect the qualitative appearance of the tide during
  periapse passage.}
\label{fig:schwpsi0_ecc}
\end{figure*}

Let us now examine tides from eccentric orbits in the black hole's
equatorial plane.  We again begin with Schwarzschild black holes, and
examine the tidal field $C_+$ for an orbit with $p = 8M$, $e = 0.5$.
The left-hand panels of Fig.\ {\ref{fig:schwpsi0_ecc}} show the
behavior of $C_+$ close to the moment that the orbit passes through
periapsis.  We examine this field in the hole's equatorial plane,
$\theta = 90^\circ$, and at four evenly spaced axial angles, $\psi =
0^\circ$, $90^\circ$, $180^\circ$, and $270^\circ$.  Note that the
cross polarization $C_\times$ vanishes for all equatorial orbits, so
we do not show it in any of our figures.  (Away from the equatorial
plane, $C_\times$ is non zero, but is qualitatively quite similar to
$C_+$.)  The right-hand panel of Fig.\ {\ref{fig:schwpsi0_ecc}} shows
the orbit's radius as a function of ingoing time near periapse
passage.

The behavior of $C_+$ at $\psi = 0^\circ$ can be regarded as a
prototype for the tidal field from eccentric orbits: there is a large
spike at roughly the same time as periapsis, with smaller scale
oscillations before and after.  These dynamics in $C_+$ occur when the
small body is closest to the event horizon.  To quantify this, we have
marked with large dots the moments at which the orbital radius is $r =
10^{1/3}r_{\rm min}$ ($v \simeq 295M$ and $v \simeq 475M$).  Since at
leading order the tide scales as $1/r^3$, we expect that the tide will
be about an order of magnitude smaller than its peak at these moments.
Comparing the left-hand panels, we see that $C_+$ in all cases is at
least a factor of ten smaller than its peak value at these times.

The remaining three left-hand panels show how the tide is modulated by
the azimuthal angle.  In all cases, we see three oscillations, but the
relative amplitude of these oscillations varies significantly with the
value of $\psi$ at which the field is measured: the middle oscillation
is ``large'' and the other two small for the prototypical form we see
at $\psi = 0^\circ$, but all three wiggles are of nearly equal
amplitude at $\psi = 180^\circ$.  As we saw in the circular case
(cf.\ Fig.\ {\ref{fig:schwpsi0_circ}}), the position at which the
field is measured significantly affects the tide.

\begin{figure*}[ht]
\includegraphics[width = 0.48\textwidth]{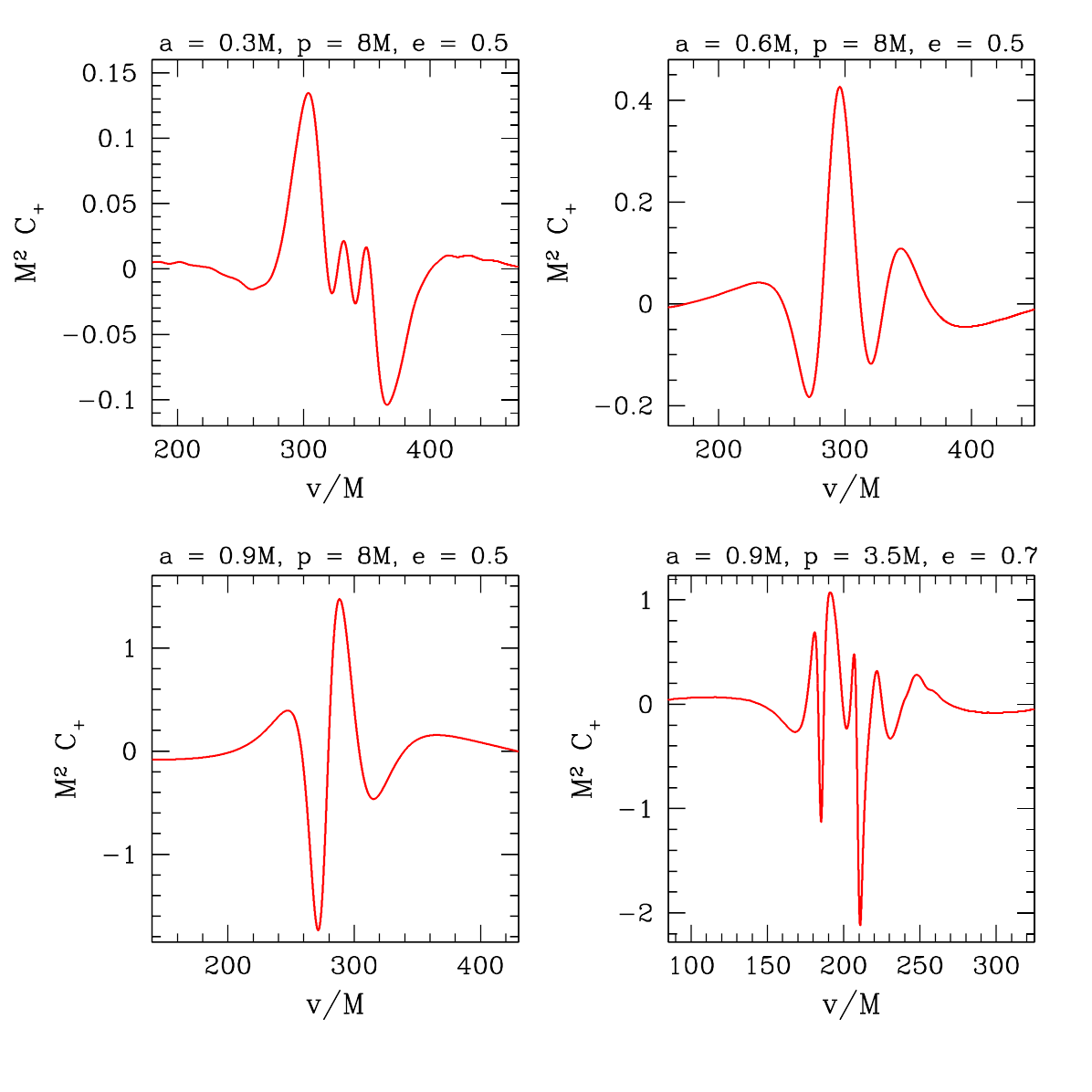}
\includegraphics[width = 0.48\textwidth]{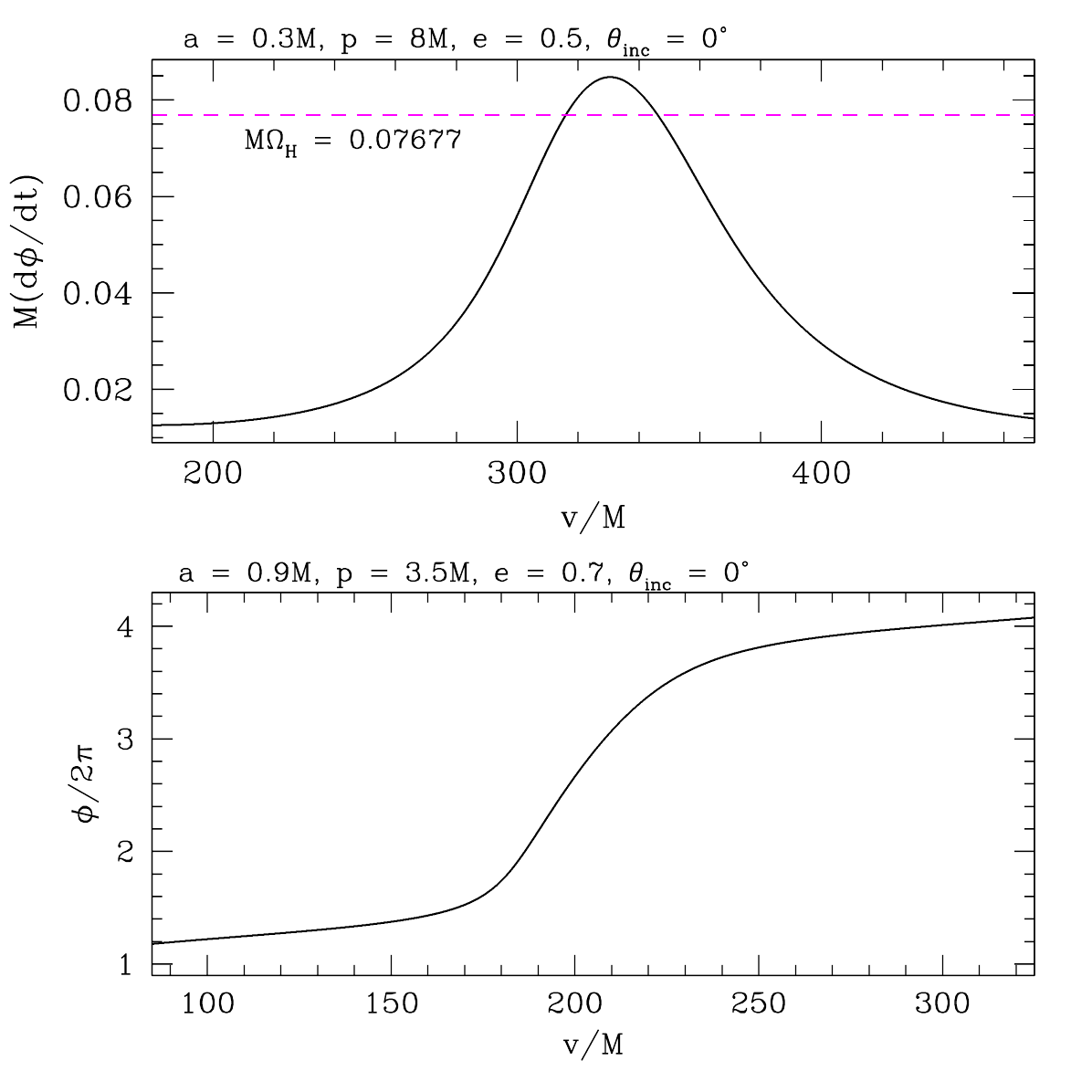}
\caption{Examples of the on-horizon tidal field $C_+$ for equatorial
  eccentric orbits of Kerr black holes (left), and features of the
  orbits which we use to explain some of the behavior we find (right).
  The cases shown here demonstrate several examples of interesting
  behavior for strong-field Kerr orbits.  On the left, the top panels
  and the bottom left panel are for orbits with $p = 8M$, $e = 0.5$,
  and show orbits about black holes with spins $a = 0.3M$, $a = 0.6M$,
  and $a = 0.9M$; the bottom right panel on the left is for an orbit
  with $p = 3.5M$, $e = 0.7$ about a black hole with $a = 0.9M$.  In
  all cases, we examine the field at $\theta = 90^\circ$, $\psi =
  0^\circ$; $C_\times$ vanishes there for all equatorial orbits.  The
  behavior we see in the top-right and bottom-left panels is very
  similar to the prototype behavior of $C_+$ we saw in
  Fig.\ {\ref{fig:schwpsi0_ecc}}: a large spike coinciding with
  periapse passage, and smaller-scale oscillations before and after.
  The behavior we see in the top-left and bottom-right deviates from
  the prototype in interesting ways.  For $a = 0.3M$, $p = 8M$, $e =
  0.5$, $C_+$ undergoes additional very small-amplitude oscillations
  during the spike.  On the right, the top panel compares the orbit's
  axial speed $d\phi/dt$ with the horizon's spin frequency
  $\Omega_{\rm H}$.  The small amplitude oscillations occur during the
  brief span in which $d\phi/dt$ exceeds $\Omega_{\rm H}$.  For $a =
  0.9M$, there are multiple high-amplitude oscillations near the
  largest spike.  This is correlated with ``whirling'' orbital
  dynamics.  As shown in the bottom panel on the right, the orbit
  wraps around the black hole multiple times at periapsis before
  ``zooming'' back out to apoapsis.}
\label{fig:kerrpsi0_ecc}
\end{figure*}

Figure {\ref{fig:kerrpsi0_ecc}} examines the tide for eccentric
equatorial Kerr black hole orbits.  We show $C_+$ for orbits with $p =
8M$, $e = 0.5$ about black holes with spins $a = 0.3M$, $a = 0.6M$,
and $a = 0.9M$, as well as an orbit with $p = 3.5M$, $e = 0.7$ about a
black hole with spin $a = 0.9M$.  Two of the examples ($p = 8M$, $e =
0.5$ for $a = 0.6M$ and $a = 0.9M$) are similar to the prototype
eccentric tide we examined for Schwarzschild (the $\psi = 0^\circ$
case of Fig.\ {\ref{fig:schwpsi0_ecc}}): a large spike near periapsis,
with smaller scale oscillations before and after.  The only notably
new feature we see in these examples is the rather different amplitude
of $C_+$ as compared to the Schwarzschild case, and as compared to
each other.  This is explained similarly to how we explained the
varying tidal amplitudes of circular Kerr orbits
(cf.\ Fig.\ {\ref{fig:kerrpsi0_circ}} and associated discussion): the
amplitude of each mode $\Lambda$ is proportional to $p_\Lambda =
\omega_\Lambda - m\Omega_{\rm H}$.  Modes can be significantly
suppressed when the orbit frequencies are close to $\Omega_{\rm H}$.

The cases $a = 0.3M$, $p = 8M$, $e = 0.5$ (top-left panel of
Fig.\ {\ref{fig:kerrpsi0_ecc}}) and $a = 0.9M$, $p = 3.5M$, $e = 0.7$
(bottom-right panel) both demonstrate significate deviations from this
prototype.  Consider the $a = 0.3M$ case first: the spike at periapse
passage in this case is interrupted by about a cycle and a half of
very small amplitude wiggle.  This phenomenon appears to arise because
of a change in the relative angular speeds of the orbit and of the
event horizon during periapse passage.

To understand this, recall (as discussed in paper I) that the tidal
field $\Psi_0$ (and hence $C_+$) vanishes for orbits that co-rotate
with the event horizon.  This only occurs for circular, equatorial
orbits, and is simple to understand: such orbits are characterized by
only one frequency, $\Omega_\phi$, and so $\omega_{mkn} \to
m\Omega_\phi$.  If $\Omega_\phi = \Omega_{\rm H}$, then $p_\Lambda =
m(\Omega_\phi - \Omega_{\rm H}) = 0$, and by Eq.\ (\ref{eq:Psi0mode})
$\Psi_0 = 0$.

In our case, the orbit does not co-rotate with the horizon for all
time, but it co-rotates at two moments as it moves through periapsis.
The top panel on the right of Fig.\ {\ref{fig:kerrpsi0_ecc}} shows
$d\phi/dt$ near periapse passage, comparing it to the horizon's spin
frequency $\Omega_{\rm H}$.  The orbit's angular speed is slower than
the horizon's spin until $v \simeq 315.9M$.  It is then faster than
the horizon until $v \simeq 346.3M$, returning to a slower angular
speed than the horizon.  The small-scale oscillations in $C_+$ occur
almost precisely during the moments that the orbit overtakes the
hole's rotation.  The tidal field oscillates with small amplitude as
the orbit passes through co-rotation and back near its periapse
passage.

The additional oscillations we see in the periapse spike for the case
$a = 0.9M$, $p = 3.5M$, $e = 0.7$ are simpler to explain.  This orbit
has a ``zoom-whirl'' structure, in which the small body ``whirls''
multiple times around the event horizon at periapsis before
``zooming'' back to apoapsis.  The bottom panel on the right of
Fig.\ {\ref{fig:kerrpsi0_ecc}} shows the number of windings about the
horizon, $\phi/2\pi$, that the orbit executes as a function of time.
The orbit winds the horizon about two and a half times in the time
interval $180 M \lesssim v \lesssim 230M$ (periapsis occurs at $v =
203M$, near the middle of this range).  The multiple oscillations in
$C_+$ occur during the period in which the orbit is whirling close to
the black hole.

\subsection{Summary of tidal behavior}

We conclude our discussion of the tide by summarizing the features
that we found above.  For the most part, we find that the dynamics of
the fields $C_{+,\times}$ correlate with the dynamics of the orbit.
For circular orbits, the tides are essentially sinusoidal, with the
amplitude modulated by the axial angle at which the field is measured.
Thanks to frame dragging, for the Kerr case this modulation becomes
associated with the orbit, leading to a dynamical modulation of the
tide's amplitude.

For eccentric orbits, the prototype form of the tide is a large spike
at periapse passage, modulated by the axial angle in a manner similar
to the axial modulation we saw with circular orbits.  ``Zoom whirl''
orbits, fairly common in the strong field of rapidly spinning black
holes, are orbits in which the orbit ``whirls'' around the black hole
at periapsis, completing multiple revolutions before ``zooming'' back
out to apoapsis.  In such cases, multiple oscillations occur during
the large amplitude periapse spike.

Certain interesting and seemingly subtle features of the tides we find
originate in the fact that modes of the tidal field are proportional
to $p_\Lambda = \omega_\Lambda - m\Omega_{\rm H}$.  This factor varies
quite a bit depending on the hole's spin and the nature of the orbit.
This leads to substantial variation in the amplitude of the tide when
we consider a sequence in which the orbit's geometry is held fixed and
the black hole spin is varied.  It also leads to interesting behavior
when we consider orbits for which $\omega_\Lambda \simeq m\Omega_{\rm
  H}$.

\begin{figure}[ht]
\includegraphics[width = 0.48\textwidth]{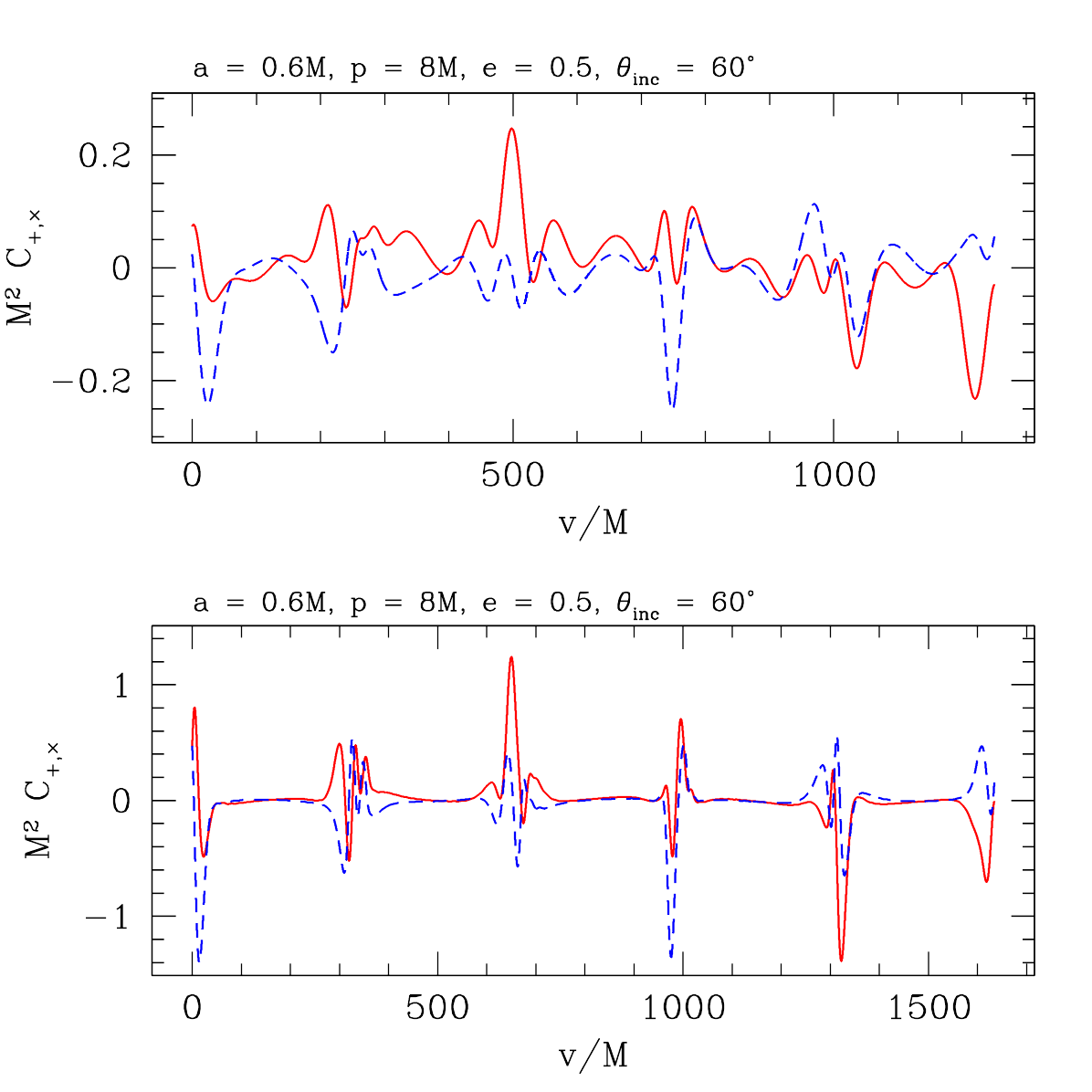}
\caption{Example of the on-horizon tidal polarizations $C_{+,\times}$
  for generic Kerr orbits.  Both panels show the tide for orbits with
  $p = 8M$, $\theta_{\rm inc} = 60^\circ$ about black holes with spin
  $a = 0.6M$; top is for an orbit with the relatively low eccentricity
  $e = 0.2$, bottom is for the much larger value $e = 0.5$.  The solid
  (red) curves show $C_+$, and the dashed (blue) curves show
  $C_\times$.  In both cases, we show the tide resulting from five
  complete radial cycles of the orbit's motion.  For $e = 0.2$, we see
  a blending of features of the circular inclined and eccentric
  equatorial tides, with larger amplitude spikes occuring near
  periapse passage, and sinusoidal oscillations between these spikes.
  For $e = 0.5$, the behavior is dominated by the very large spike
  seen at periapse passage, and is quite similar in form to the
  eccentric equatorial case.}
\label{fig:generic_tide}
\end{figure}

Figure {\ref{fig:generic_tide}} shows how the features of tides from
inclined circular and equatorial eccentric cases combine for generic
orbits.  We examine two orbits with $p = 8M$, $\theta_{\rm inc} =
60^\circ$ about a black hole spin $a = 0.6M$.  One orbit has
eccentricity $e = 0.2$, the other $e = 0.5$.  The low eccentricity
case blends the features of the inclined circular and equatorial
eccentric limits in fairly straightforward way: we find relatively
high amplitude tidal spikes at each periapse passage, with sinusoidal
tidal oscillations between each passage.  In the case $e = 0.5$, the
behavior we see practically cannot be distinguished from an equatorial
eccentric case.  The eccentricity is sufficiently large in this case
that there is a very large contrast between the tidal spike near
periapsis and the much weaker tide at apoapsis.  Any oscillations
between periapse passages are dwarfed by the much more important spike
in the tide when the orbit is closest to the black hole.

\section{Horizon dynamics I: Consistency test for the Schwarzschild limit}
\label{sec:resultsI}

We now turn to our examination of the dynamics of the horizon's
geometry.  We begin by first testing whether the Schwarzschild limit
exhibits the correct behavior.  These black holes are spherically
symmetric, so there is no physical distinction between an equatorial
orbit ($\theta_{\rm inc} = 0^\circ$) and an orbit of arbitrary
inclination.  Our representation of these orbits will certainly be
different, but this is due to the coordinate orientation we have
chosen.  (By contrast, when $a \ne 0$, the black hole's spin axis
picks out a preferred spatial direction.)  We thus expect that many
properties related to black hole perturbations should become invariant
with respect to orbit inclination for $a = 0$, or else vary in a
simple way.

This limiting behavior has been discussed in past work, in particular
describing how the amplitude of gravitational waves and the energy
that they carry varies as the orbit's inclination varies.  As one
example, consider the energy carried by gravitational waves.  The
total energy carried by a given $l$-mode must be constant as a
function of orbital inclination:
\begin{eqnarray}
\left(\frac{dE}{dt}\right)_l &\equiv&
\sum_{mkn}\left(\frac{dE}{dt}\right)_{lmkn}
\nonumber\\
&=& \mbox{constant with $\theta_{\rm inc}$}\;.
\label{eq:dEdt_l}
\end{eqnarray}
The sum in Eq.\ (\ref{eq:dEdt_l}) is taken over $m$ from $-l$ to $l$,
and over $n$ from $-\infty$ to $\infty$.  The sum over $k$ in
principle runs from $-\infty$ to $\infty$, though many modes do not
actually contribute, as we discuss momentarily.

Although the summed flux $(dE/dt)_l$ does not vary with $\theta_{\rm
  inc}$, the distribution of gravitational-wave power among the
harmonic indices varies with inclination quite a bit.  In the
Schwarzschild limit, $\Omega_\theta = \Omega_\phi$.  Consider two
orbits which are identical except for inclination.  One is equatorial
($\theta_{\rm inc} = 0^\circ$), the other is not.  Power in an axial
$m$-mode at $\theta_{\rm inc} = 0^\circ$ becomes distributed among
polar $k$-modes and axial modes with $m' = (m - k)$ in the inclined
orbit.  The way in which the power is so distributed is easily deduced
from the rotation properties of spherical harmonics:
\begin{equation}
\frac{(dE/dt)_{l(m-k)kn}(\theta_{\rm inc})}{(dE/dt)_{lm0n}(\theta_{\rm
    inc} = 0^\circ)} = \left|{\cal D}^l_{(m-k)m}(\theta_{\rm
  inc})\right|^2\;.
\end{equation}
Here, ${\cal D}^l_{(m-k)m}$ is a Wigner function, which relates the
spherical harmonic $Y_{lm}$ at $\theta$ to the harmonic $Y_{l(m-k)}$
at $\theta - \theta_{\rm inc}$.  (This relation implies that there is
no power in any mode with $|m + k| \ge l$.)  Further discussion of
this relation is given in Refs.\ {\cite{h00}} (with a few minor
errors) and {\cite{dh06}} (which corrects those errors).

What applies to the gravitational wave flux likewise applies to all
the quantities which describe tidal distortions of a Schwarzschild
black hole's event horizon.  We find that, in all cases we have
checked, quantities transform under rotation exactly as they should.
This is not terribly surprising, since this property of our code has
been checked very carefully in previous analyses.  It is reassuring,
however, that the modifications we have made to compute the horizon's
tidal distortion have not broken this behavior.

\begin{figure}[ht]
\includegraphics[width = 0.48\textwidth]{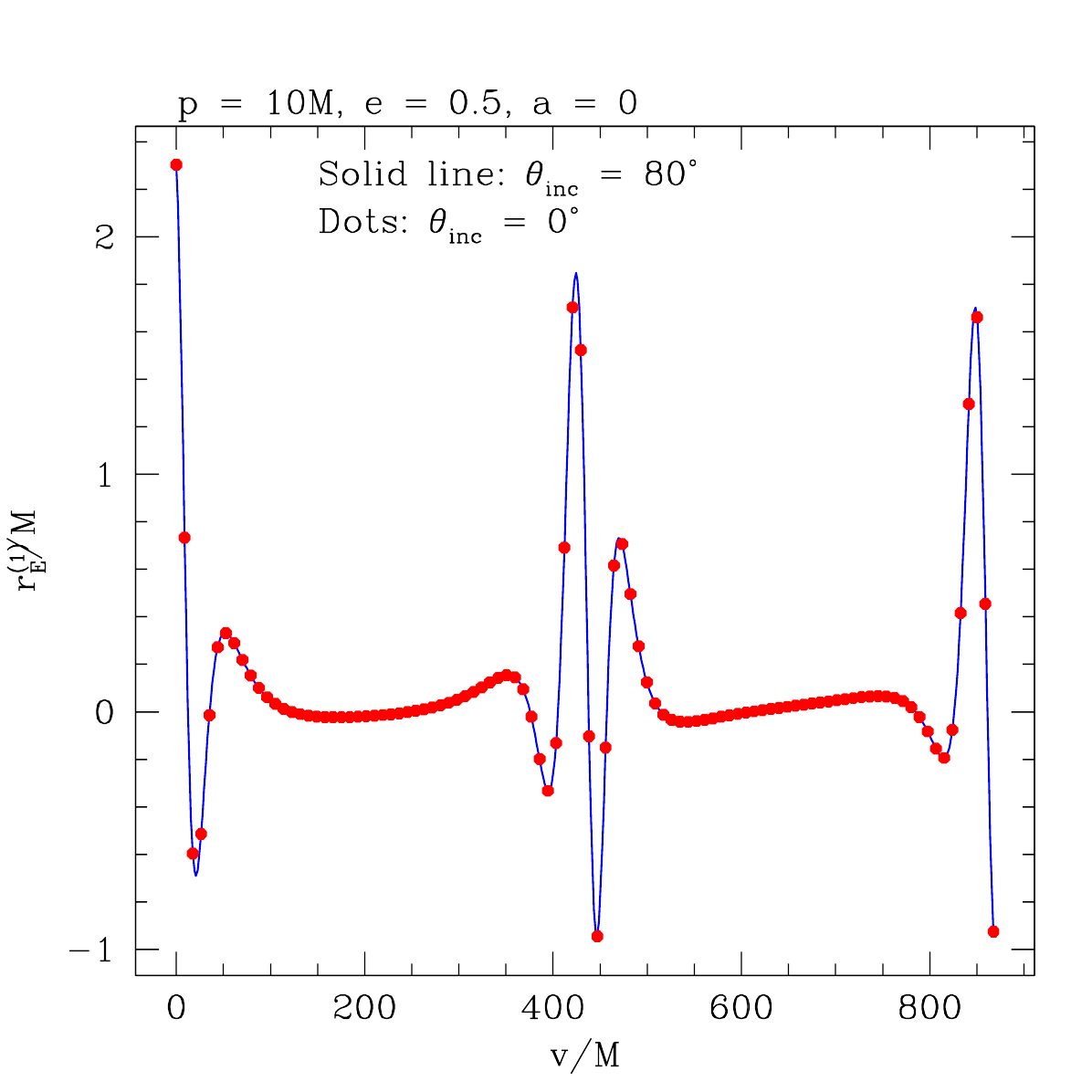}
\caption{A test for consistency of our results in the $a = 0$ limit:
  the distortion to the embedded horizon arising from a highly
  inclined orbit [$\theta_{\rm inc} = 80^\circ$; solid (blue) curve]
  as measured by an observer sitting $\theta_{\rm inc}$ above the
  equator, and from an equatorial orbit [$\theta_{\rm inc} = 0^\circ$;
    points (red)] as measured by an observer on the equator.  Both
  data sets include modes to $l = 7$; we estimate that contributions
  beyond this affect the horizon's shape at a level $r^{(1)}_{\rm
    E}/2M \lesssim 10^{-7}$.  The two data sets agree to within
  numerical accuracy, as they should --- a Schwarzschild black hole is
  spherically symmetric, so there is no unique notion of the hole's
  equator.}
\label{fig:schwtest}
\end{figure}

Figure {\ref{fig:schwtest}} shows one example of a test for the
rotational consistency of Schwarzschild horizon distortions.  Consider
two orbits around a Schwarzschild black hole, both with $p/M = 10$ and
$e = 0.5$.  One orbit is equatorial in the coordinates we impose, the
other is highly inclined ($\theta_{\rm inc} = 80^\circ$) in these
coordinates.  Due to spherical symmetry, the horizon distortion for
the equatorial case should be identical to the horizon distortion in
the inclined case, correcting for the tilt of $\theta_{\rm inc}$.

In this figure, we show the perturbation to the radius of the
horizon's embedding surface, $r^{(1)}_{\rm E}$, for these two cases.
The solid (blue) line shows the distortion for the inclined case as
measured at $\psi = 0$, $\theta = 10^\circ$ (i.e., rotated
$\theta_{\rm inc}$ from the equator).  The dots (red) show the
distortion at $\psi = 0 $ on the equator for the equatorial orbit.  We
include all modes which contribute to the horizon's distortion up to
$l = 7$; we estimate that modes beyond this affect the horizon's shape
at a level $r^{(1)}_{\rm E}/2M \alt 10^{-7}$.

Although these calculations were done using very different orbits, and
very different modes enter the expansion, the horizon distortions
$r^{(1)}_{\rm E}$ we find are essentially identical, only differing
due to accumulated round-off error at a level $\lesssim\epsilon$,
where $\epsilon \simeq 10^{-10}$ is a parameter controlling the
accuracy of numerical integrals.  If both curves had been plotted as
solid lines, they would have been indistinguishible here.  This is a
typical example of how our code handles this consistency test.

\section{Horizon dynamics II: Applied tidal field and resulting shear}
\label{sec:resultsII}

We begin with an analysis of the horizon's response to an equatorial,
eccentric orbit.  The applied tidal field varies from quite strong
near periapsis [$r = r_{\rm min} = p/(1 + e)$] to weak near apoapsis
[$r = r_{\rm max} = p/(1 - e)$], giving us a chance to study the
horizon's response for a wide range of applied tidal field.

\subsection{Relative phase of the tide and shear}
\label{sec:tideshearphase}

We begin our study of the shear induced on the horizon by examining
its phase relative to the driving tide.  In paper I, the driving tide
was stationary, and the difference between the tide and the response
amounted to a simple phase shift.  For generic orbits, the difference
is not so simple.

\begin{figure}
\includegraphics[width = 0.48\textwidth]{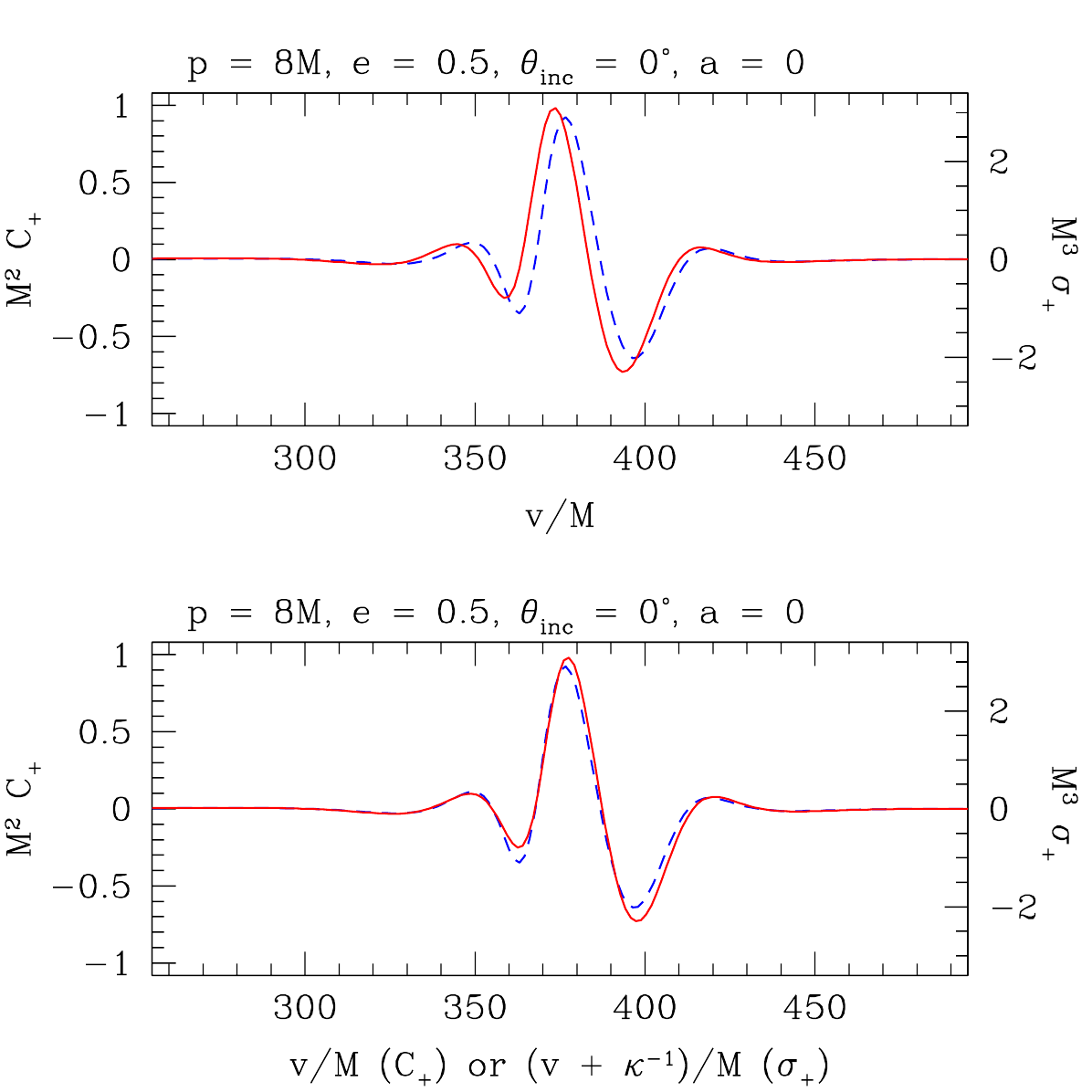}
\caption{Applied tidal field versus shear response for an eccentric
  equatorial orbit of a non-spinning black hole.  Data are for an
  orbit with $p = 8M$, $e = 0.5$, $\theta_{\rm inc} = 0^\circ$.  We
  show the on-horizon Weyl curvature polarization $C_+$ [dashed (blue)
    curves], as well as the resulting shear polarization $\sigma_+$
  [solid (red) curves].  Both fields are plotted at $\theta =
  90^\circ$, $\psi = 0^\circ$, and include all modes up to $l = 9$.
  The top panel shows $C_+$ and $\sigma_+$ as functions of ingoing
  time $v$.  Notice that $\sigma_+$ appears to lead $C_+$ by an almost
  constant time interval.  The bottom panel shows the same data, but
  with $\sigma_+$ shifted by $\Delta v = \kappa^{-1} = 4M$.  The shear
  response lines up almost perfectly with the driving tide in this
  panel, showing that the shear $\sigma_+$ leads the tide by
  $\kappa^{-1}$ in the Schwarzschild limit.}
\label{fig:a0.0shearvstide}
\end{figure}

\begin{figure}
\includegraphics[width = 0.48\textwidth]{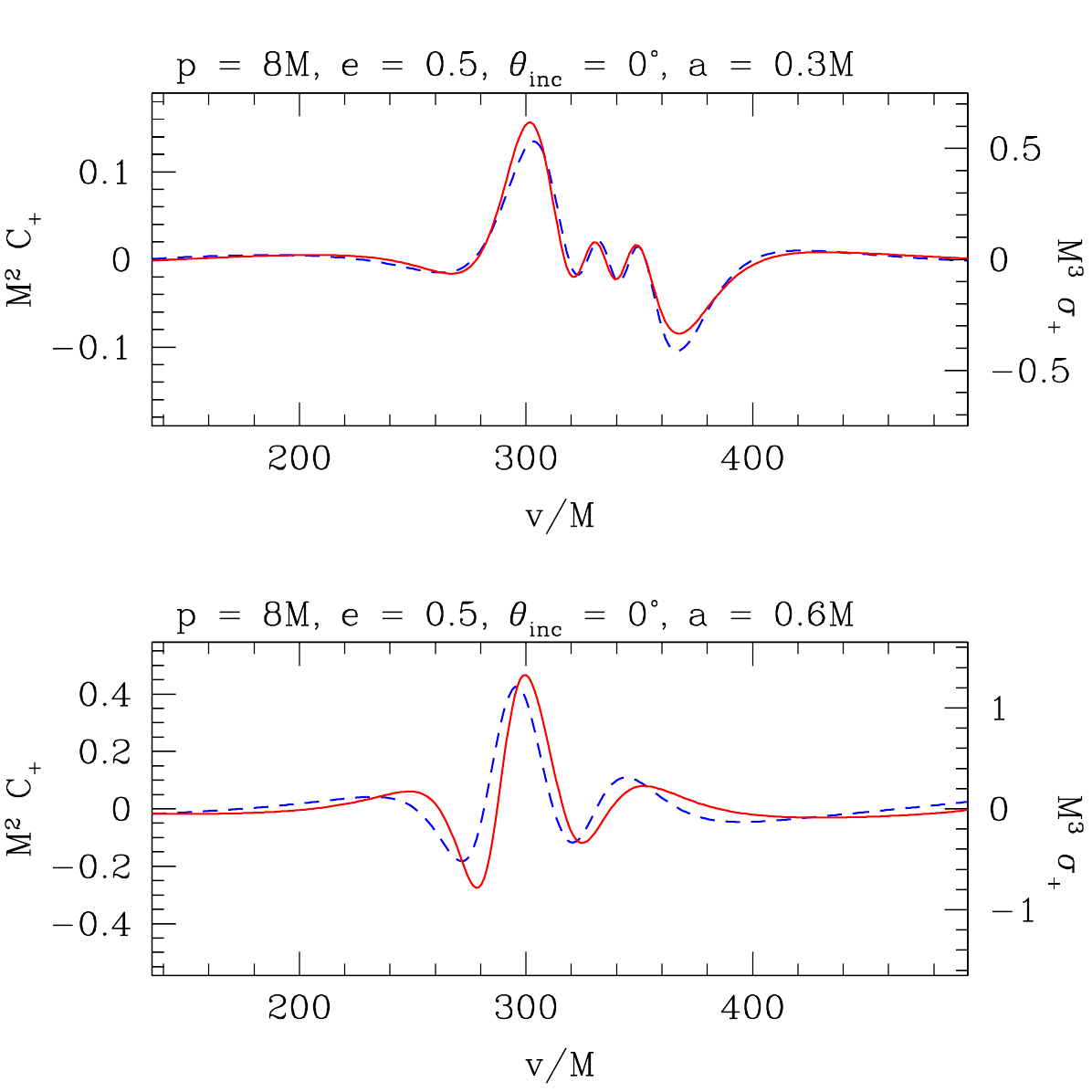}
\caption{Applied tidal field versus shear response for eccentric
  equatorial orbits of spinning black holes.  Data in both panels are
  for an orbit with $p = 8M$, $e = 0.5$, $\theta_{\rm inc} = 0^\circ$,
  and include modes up to $l = 9$.  Top is for a hole with spin $a =
  0.3M$, bottom for $a = 0.6M$.  As in
  Fig.\ {\ref{fig:a0.0shearvstide}}, we show the on-horizon Weyl
  curvature polarization $C_+$ [dashed (blue) curve] and the shear
  polarization $\sigma_+$ that results [solid (red) curve].  All data
  are plotted at $\theta = 90^\circ$, $\psi = 0^\circ$.}
\label{fig:a0.3_0.6shearvstide}
\end{figure}

\begin{figure}
\includegraphics[width = 0.48\textwidth]{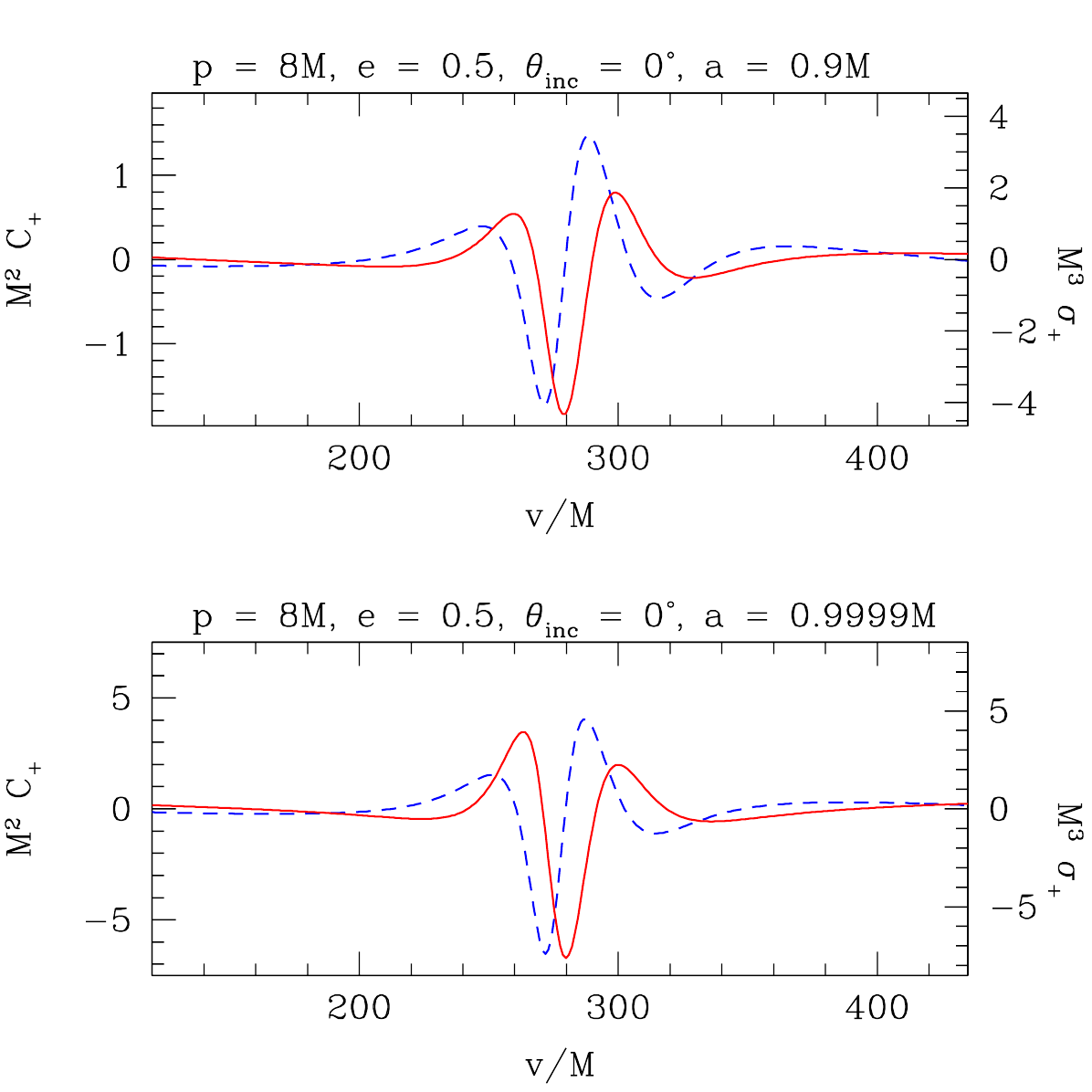}
\caption{Applied tidal field versus shear response for eccentric
  equatorial orbits of spinning black holes.  Data in both panels are
  for an orbit with $p = 8M$, $e = 0.5$, $\theta_{\rm inc} = 0^\circ$,
  and include modes up to $l = 9$.  Top is for a hole with spin $a =
  0.9M$, bottom for $a = 0.9999M$.  As in
  Figs.\ {\ref{fig:a0.0shearvstide}} and
  {\ref{fig:a0.3_0.6shearvstide}}, we show the on-horizon Weyl
  curvature polarization $C_+$ [dashed (blue) curves] and the shear
  polarization $\sigma_+$ that results [solid (red) curves].  All data
  are plotted at $\theta = 90^\circ$, $\psi = 0^\circ$.}
\label{fig:a0.9_0.9999shearvstide}
\end{figure}

Figures {\ref{fig:a0.0shearvstide}}, {\ref{fig:a0.3_0.6shearvstide}},
and {\ref{fig:a0.9_0.9999shearvstide}} compare the tidal field and
shear in five different situations.  In all cases, the orbit has $p =
8M$, $e = 0.5$, $\theta_{\rm inc} = 0^\circ$, but the black hole spin
varies over $a/M \in [0,0.3,0.6,0.9,0.9999]$.  We include all modes up
to $l = 9$ in these plots.  We compare one polarization of the
on-horizon Weyl tensor, $C_+$ [dashed (blue) curves], to the
corresponding polarization of the horizon's shear, $\sigma_+$ [solid
  (red) curves].  The orbits are all equatorial, so we examine these
quantities in the holes' equatorial planes: all data are shown at the
point $\theta = 90^\circ$, $\psi = 0^\circ$.

Begin with Fig.\ {\ref{fig:a0.0shearvstide}}, which shows $C_+$ versus
$\sigma_+$ for orbits of a Schwarzschild black hole.  The top panel of
this figure shows that the horizon's response leads the driving tide
by what is apparently a constant offset.  To understand this, consider
again Eq.\ (\ref{eq:modalphaseoffset}):
\begin{equation}
\frac{\sigma_\Lambda}{\Psi_{0,\Lambda}} =
\frac{\exp{\left[-i\arctan(p_\Lambda/\kappa)\right]}}{\sqrt{p_\Lambda^2
    + \kappa^2}}\;.
\end{equation}
In the Schwarzschild limit, $p_\Lambda = \omega_\Lambda$, and
$\kappa^{-1} = 4M$.  Each mode $\sigma_\Lambda$ of the shear response
leads the driving tide $\Psi_0$ by $4M\omega_\Lambda$ radians.  This
is equivalent to $\sigma$ leading $\Psi_0$ in time by $4M$.  We check
this in the bottom panel of Fig.\ {\ref{fig:a0.0shearvstide}}: this
plot is identical to the top panel of
Fig.\ {\ref{fig:a0.0shearvstide}}, but we have shifted $\sigma_+$ by
$\Delta v = \kappa^{-1} = 4M$.  Notice that the tide and the shear are
almost precisely aligned in this panel, confirming that the responses
here differ primarily by a temporal offset of $\kappa^{-1} = 4M$.

As the black hole's spin increases, the shift between the applied tide
and the shear response becomes more complicated: the timescale
$\kappa^{-1}$ becomes larger as $a \to M$, and the wavenumber
$p_\Lambda = \omega_\Lambda - m\Omega_{\rm H}$ which enters the mode
ratio (\ref{eq:modalphaseoffset}) differs significantly from the
frequency $\omega_\Lambda$.  We can see the impact of this change in
Fig.\ {\ref{fig:a0.3_0.6shearvstide}}.  In the top panel, we examine
$C_+$ and $\sigma_+$ for the same orbit used in
Fig.\ {\ref{fig:a0.0shearvstide}} ($p = 8M$, $e = 0.5$, $\theta_{\rm
  inc} = 0^\circ$), but now about a black hole with spin $a = 0.3M$.
In this case, the tide and the shear are nearly coincident as a
function of ingoing time $v$ (including the small-amplitude
oscillations in the periapse spike we discussed in
Sec.\ {\ref{sec:tide}}).  In the bottom panel, we plot $C_+$ and
$\sigma_+$ for this orbit about a black hole with spin $a = 0.6M$.
The tide now leads the shear, and the shapes are not congruent.
Empirically, we find that if we shift $\sigma_+$ by $\delta v \simeq
3.8M$ we can make the largest peaks line up.  Other features, however,
do not line up so well; the differing behaviors of $C_+$ and
$\sigma_+$ cannot be ascribed to a simple time shift.

The trend seen in Fig.\ {\ref{fig:a0.3_0.6shearvstide}} continues in
Fig.\ {\ref{fig:a0.9_0.9999shearvstide}}, which shows $C_+$ and
$\sigma_+$ for the same orbit about black holes with $a = 0.9M$ (top)
and $a = 0.9999M$ (bottom).  We again see that the tidal field $C_+$
leads the shear response $\sigma_+$.  We can match the largest peaks
by shifting $\sigma_+$ by $\delta v\simeq 8M$ in the case $a = 0.9M$,
and by $\delta v\simeq 9M$ in the case $a = 0.9999M$.  However, none
of the other features align when we do this, indicating that the shift
at these large spins cannot be described as a simple shift in time.

One interesting feature that comes across as we review
Figs.\ {\ref{fig:a0.0shearvstide}} --
{\ref{fig:a0.9_0.9999shearvstide}} is the transition from shear
leading the tide at $a = 0$ to shear lagging the tide for $a > 0.3M$.
This transition is reminiscent of the behavior of the tidal bulge that
was seen in paper I.  There, we found for circular equatorial orbits
that the tidal bulge leads the applied tide at small spin, and lags
the applied tide at large spin.  At least in the small $a$ and large
$a$ limits, this could be understood in the circular equatorial case
as reflecting the relative angular frequencies of the orbit and the
black hole.  Qualitatively similar behavior clearly shows up for these
dynamical situations, although quantifying it is not so
straightforward since these orbits have a more complicated
time-frequency structure.

\subsection{High-frequency oscillations at high spin}
\label{sec:highspinwiggles}

\begin{figure*}
\includegraphics[width = 0.48\textwidth]{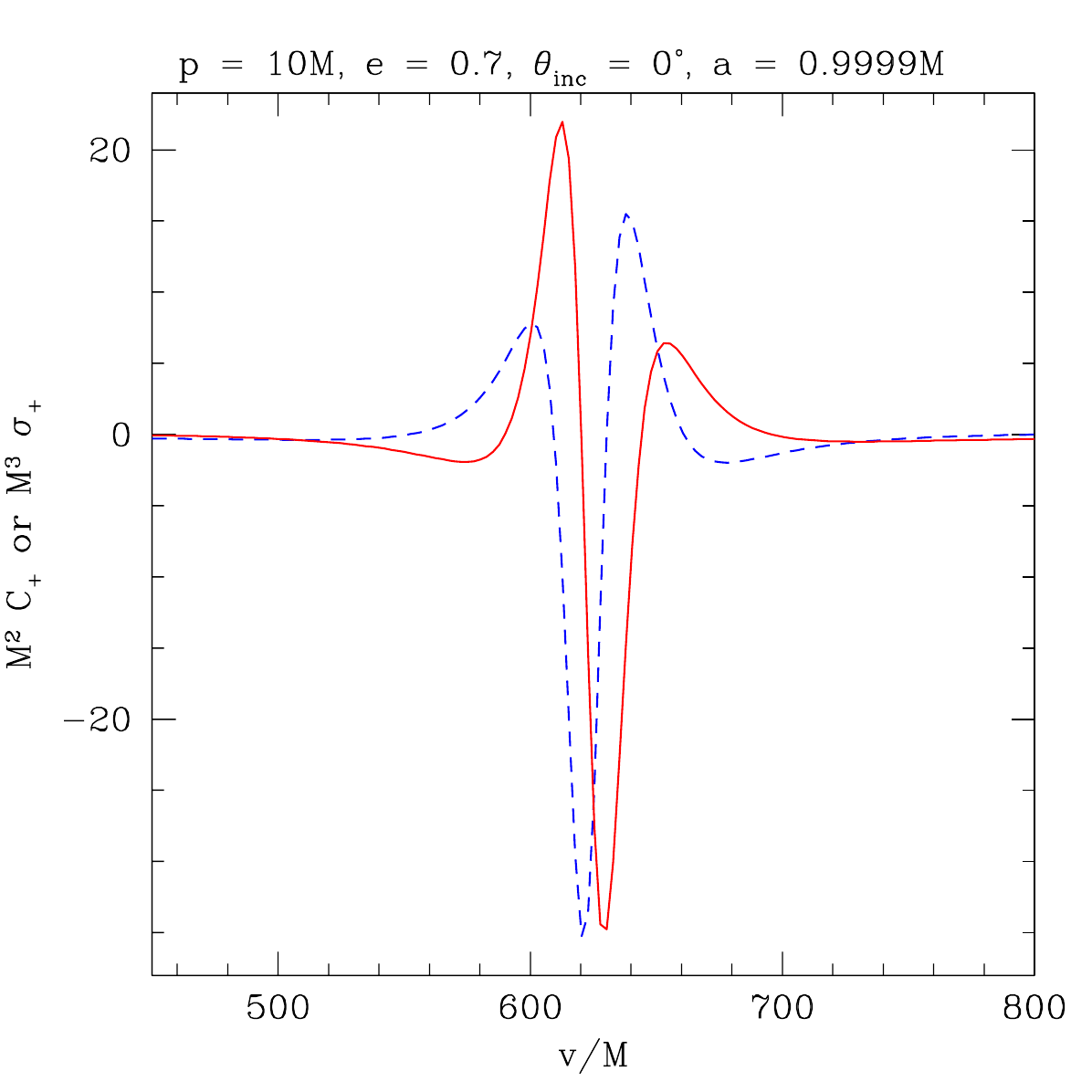}
\includegraphics[width = 0.48\textwidth]{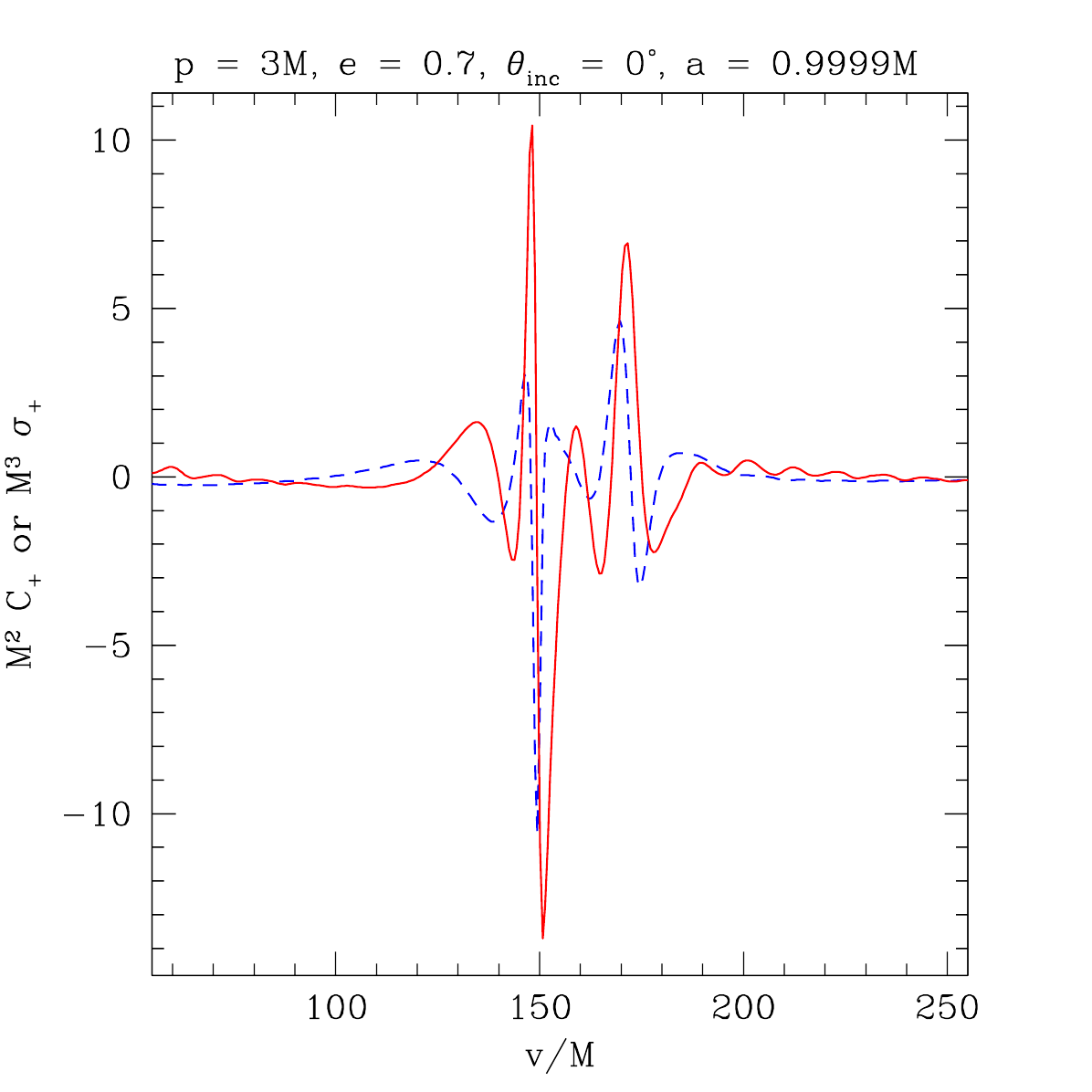}
\caption{Detailed study of the horizon's shear response $\sigma_+$
  [solid (red) curves] given a driving tidal field $C_+$ [dashed
    (blue) curves] for the case of very rapid spin, $a = 0.9999M$.  In
  both panels, we consider equatorial ($\theta_{\rm inc} = 0^\circ$)
  orbits with eccentricity $e = 0.7$, and include all modes up to $l =
  9$.  In the left-hand panel, the orbit has semi-latus rectum $p =
  10M$; in the right, $p = 3M$.  For the case $p = 10M$, no
  particularly noteworthy feature is evident.  This behavior is
  qualtitatively similar to the shear response that we see across a
  wide range of orbits: $\sigma$ has a shape similar to $C$, but is
  offset as a function of time.  However, for high spins and
  strong-field orbits, new behavior emerges: low-amplitude,
  high-frequency wiggles can be seen in the shear between the
  high-amplitudes ``bursts'' corresponding to the orbit's periapse
  passage.  Notice that these wiggles are not present in the tidal
  field, only in the shear response.}
\label{fig:a0.9999_compare}
\end{figure*}

At the highest spins we have examined, a new phenomenon emerges: a
high-frequency oscillation in the shear $\sigma$ following the orbit's
passage through periapsis.  This oscillation decays over a time of
about $70M$ for the orbits we have examined.  An example is shown in
Fig.\ {\ref{fig:a0.9999_compare}}.  Both panels of this figure show
data for equatorial orbits with $e = 0.7$ about black holes with spin
$a = 0.9999M$.  The left panel shows data from an orbit with $p =
10M$.  The behavior of $C_+$ and $\sigma_+$ is quite similar to the
cases discussed previously: $C_+$ shows a large spike near periapse
passage, with small scale oscillations before and after; $\sigma_+$
has a similar shape, offset somewhat in time.  (Because this is an
equatorial orbit and we examine these fields on the equator,
$C_\times$ and $\sigma_\times$ are both zero.)

On the right, we show a much stronger field orbit, $p = 3M$.  The
spike at periapse passage has several large amplitude oscillations
characteristic of the ``whirling'' near periapsis common for large
spin, strong-field orbits.  This is essentially the same phenomemon
seen in the bottom-right panel on the left-hand side of
Fig.\ {\ref{fig:kerrpsi0_ecc}}.  The new phenomenon to which we call
attention are the low-amplitude, high-frequency wiggles that follow
periapse passage.  We see roughly seven low-amplitude cycles in
$\sigma_+$ between periapse spikes, decaying in amplitude as the
system evolves from one spike to the next.  These wiggles are only
apparent in the shear $\sigma_+$; we have not seen evidence of them in
the tidal field $C_+$.

As we complete this analysis, the origin of these low-amplitude
oscillations is a mystery.  They do not appear to be a numerical
artifact; we are confident that our harmonic expansion has converged,
as including additional modes does not change our results beyond the
ninth or tenth digit.  The fact that these oscillations only appear in
the shear $\sigma_+$ and not in the tidal field $C_+$ or $\Psi_0$
indicates that they cannot be related to the hole's quasi-normal
modes.  As we'll discuss in a moment, the behavior of their decay also
argues against such an explanation.  In an earlier version of this
paper, we argued that these wiggles could be understood as an imprint
of the teleological Green's function discussed in
Sec.\ {\ref{sec:horizongeom}}, with the oscillation frequency related
to the horizon's spin frequency $\Omega_{\rm H}$, and the decay to the
Green's function's decay time $\kappa^{-1} = 2r_+/\sqrt{1 - (a/M)^2}$.
On deeper analysis (prompted by our original submission's referee
report), we have concluded that the data does not support this
explanation either.

\begin{figure}
\includegraphics[width = 0.48\textwidth]{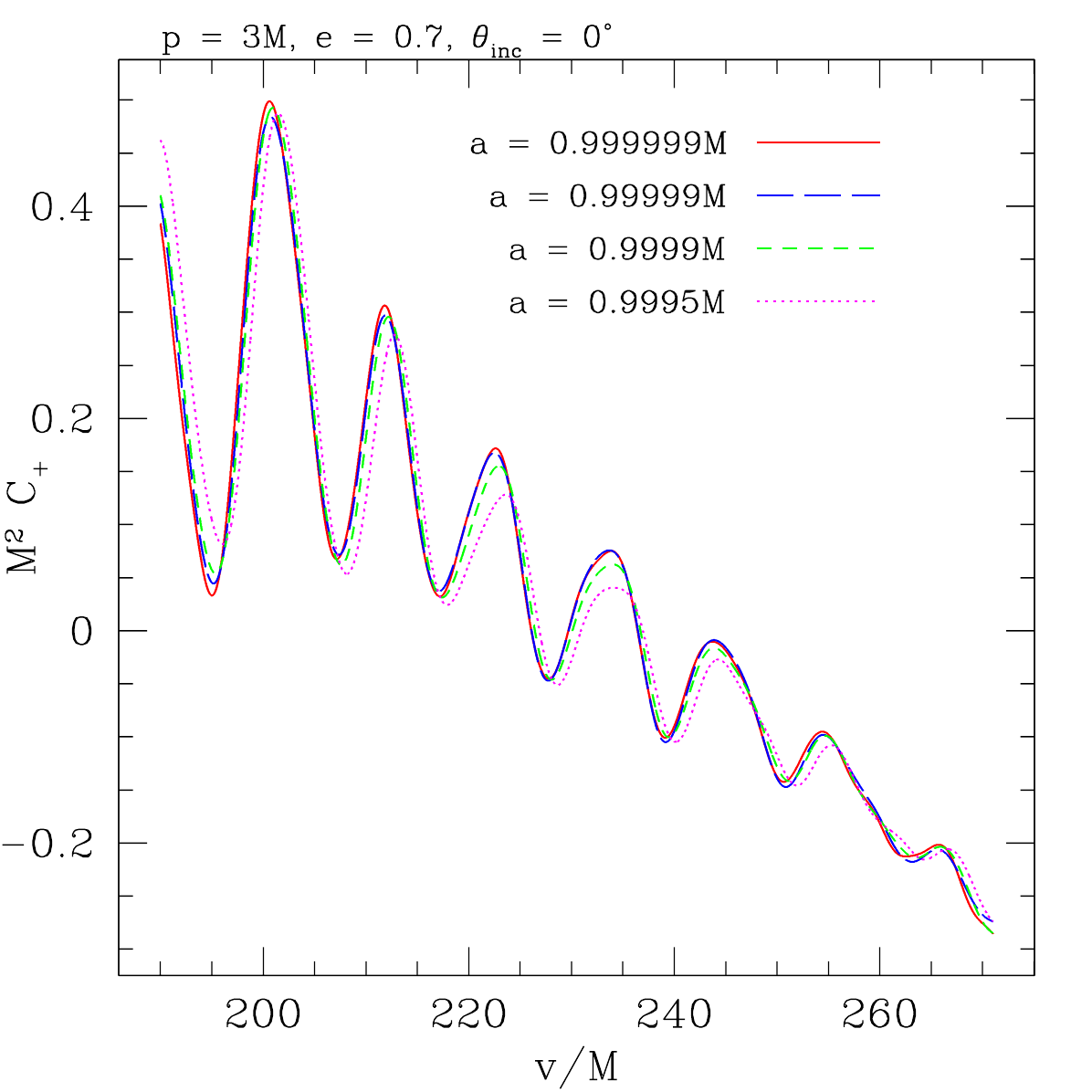}
\caption{A zoom on the high-frequency wiggles shown in the right-hand
  panel of Fig.\ {\ref{fig:a0.9999_compare}}.  Along with the $a =
  0.9999M$ case shown previously, we include data describing wiggles
  for orbits with the same values of $p$, $e$, and $\theta_{\rm inc}$,
  but for spins $a = 0.9995M$, $a = 0.99999M$, and $a = 0.999999M$.
  As described in the text, the frequency associated with the wiggles
  (which we have not succeeded in relating to other frequencies in the
  problem) decreases slightly as a function of spin.  The rate at
  which the amplitude decays does not vary significantly with spin.
  We have not been able to come up with a satisfactory explanation for
  the phenomenon of these wiggles.}
\label{fig:wigglezoom}
\end{figure}

For now, rather than offering any hypotheses attempting to explain
these wiggles, we simply lay out the empirical situation as it stands.
Figure {\ref{fig:wigglezoom}} zooms in on the region of
Fig.\ {\ref{fig:a0.9999_compare}} in which this phenomenon is
apparent.  We also include data for additional spin values for orbits
with this geometry.  The trend we see is that the frequency of the
oscillations is nearly the same for these cases, evolving slightly as
the spin moves toward the extremal limit.  The decay time likewise is
nearly constant over this range of spin.  The near constancy of the
decay time appears to rule out an explanation for these wiggles based
on black hole quasi-normal modes, or on the Green's function
(\ref{eq:sheargreens}).  Both of these explanations would require the
decay time to become dramatically longer as $a \to M$.  The form of
the teleological Green's function, for example, would lead us to
expect the decay time to increase by a factor of ten as the spin
changes from $a = 0.9999M$ to $a = 0.999999M$.  The lack of such
increase points to some other mechanism.

By fitting and subtracting a quadratic to remove the secular trend in
the data shown in Fig.\ {\ref{fig:wigglezoom}}, we can more accurately
locate the position of the peaks and estimate the frequency associated
with the wiggles.  Assuming that the phase of these wiggles follows
the form
\begin{equation}
\Phi_{\rm wiggle} = \Omega_{\rm wiggle} v + \delta\Phi\;
\end{equation}
and requiring that the phase of the peaks lie on the line
\begin{equation}
\Phi_{\rm wiggle}({\rm peak}) = 2\pi n
\end{equation}
with $n$ integer, we can estimate the wiggle frequency that best fits
our data.  The results are summarized in Table
{\ref{tab:wigglefreqs}}.  We present the best-fit frequencies versus
both the spin $a$ and a parameter $\epsilon \equiv \sqrt{1 -
  (a/M)^2}$, which characterizes the deviation of the spacetime from
extremal Kerr.

\begin{table}
\begin{tabular}{|c|c|c|c|c|c|}
\hline
$a/M$ & $\sqrt{1 - (a/M)^2}$ & $M\Omega_{\rm wiggle}$ & $M\Omega_{\rm H}$
& $M\Omega_\phi$ & $M\Omega_r$ \\
\hline
$0.9995$ & $0.03162$ & $0.5826$ & $0.4844$ & $0.0988$ & $0.0399$ \cr
\hline
$0.9999$ & $0.01414$ & $0.5805$ & $0.4930$ & $0.0988$ & $0.0399$ \cr
\hline
$0.99999$ & $0.004572$ & $0.5792$ & $0.4978$ & $0.0988$ & $0.0399$ \cr
\hline
$0.999999$ & $0.001414$ & $0.5787$ & $0.4993$ & $0.0988$ & $0.0399$ \cr
\hline
\end{tabular}
\caption{The best-fit frequency $\Omega_{\rm wiggle}$ characterizing
  the high-frequency oscillations shown in
  Fig.\ {\ref{fig:wigglezoom}}.  We present this frequency as a
  function of both the black hole spin $a/M$, and the parameter
  $\epsilon \equiv \sqrt{1 - (a/M)^2}$ which characterizes deviation
  from extremality.  For comparison, we show the hole's spin frequency
  $\Omega_{\rm H}$ for these spins, and the orbit's geodesic
  frequencies $\Omega_\phi$ and $\Omega_r$.  We see no obvious
  physically motivated way to connect $\Omega_{\rm wiggle}$ to these
  other frequencies.}
\label{tab:wigglefreqs}
\end{table}

We find that a linear fit in $\epsilon$ describes our best fit
frequencies quite well.  Performing a least-squares fit of our data,
we find
\begin{eqnarray}
M\Omega_{\rm wiggle} = (0.5786 \pm 0.0001)
+ (0.1288 \pm 0.0032)\,\epsilon\;.
\nonumber\\
\end{eqnarray}
Although the goodness of this fit with $\epsilon$ is intriguing, we
cannot yet claim any understanding for what this might signify.  In
Table {\ref{tab:wigglefreqs}}, we include the geodesic frequencies
$\Omega_\phi$ and $\Omega_r$ for the orbits used in
Fig.\ {\ref{fig:wigglezoom}}, as well as the horizon frequency
$\Omega_{\rm H}$.  As an exercise in arithmetic, we can find no
combination of these frequencies that produces the values we find for
$\Omega_{\rm wiggle}$.  Absent any compelling physical model, any
combination we did find would arguably be no more useful than
numerology.

For now, we leave this phenomenon as an intriguing empirical finding
of this analysis, and hope that additional work may explain it in the
future.

\section{Horizon dynamics III: Horizon embeddings}
\label{sec:resultsIII}

In this section, we examine dynamical horizon embeddings for several
representative orbits.  Our goal will be to show how the horizon
behaves as a function of the orbit's behavior, so we will show a
sequence of figures that show both the tidally distorted horizon and
the smaller member of the binary.  As discussed at length in paper I,
there is substantial ambiguity in such a plot, associated with the
fact that the horizon and the orbit are at different positions in a
curved spacetime.  Comparing the horizon and the orbit requires that
we carefully define exactly what is shown.  Following the choices that
we made in paper I, our plots are all shown on a slice of constant
ingoing time $v$; this is equivalent to what we called the
``instantaneous map'' in paper I.

One might wonder why, given this ambiguity, we choose to present our
data using these embedding diagrams.  Indeed, there are multiple ways
that one can present data representing the geometry of distorted black
holes.  For example, one could make a color map representing the
scalar curvature $R^{(1)}_{\rm H}$, or a color map representing the
phase between the on-horizon tide $\Psi_0$ and the resulting shear
$\sigma$.  (Note that both $\Psi_0$ and $\sigma$ are spin-weight 2
quantities, and so cannot be simply represented on a surface --- both,
for example, are multiply valued at the poles, $\theta = 0$ and
$\theta = \pi$.)  Such representations have the advantage of
presenting quantities that are less ambiguious.

In the end, we have chosen to use embeddings primarily for aesthetic
reasons.  One of our goals was to develop graphics which demonstrate
the extent to which a black hole's shape is distorted by tides from
its companion.  Although one must be careful in interpreting this
shape, embeddings provide a compelling picture of this tidal shape
distortion.  We supplement the shape with a color map which codes the
horizon's distortion from the shape it would have in the absence of a
binary companion.  In all of the figures which follow, surfaces
colored green are essentially undistorted from the embedding of an
isolated black hole; those colored red have larger radius than that of
an isolated black hole; and those colored blue have smaller radius.
We find that color maps of other quantities, such as the scalar
curvature $R^{(1)}_{\rm H}$ are visually quite similar to the color
maps we associate with the embedding.  For our purposes, horizon
embeddings, though somewhat arbitrary, convey exactly the information
that we hoped to present.  Using the tools we have developed here and
in paper I, it is straighforward to modify this analysis to focus upon
other measures of horizon distortion.

As in paper I, a major shortcoming of our use of embeddings is that we
embed the horizon in a Euclidean three-dimensional space.  This means
we are confined to spin parameter $a/M \le \sqrt{3}/2$; for faster
spins, even an undistorted horizon cannot be embedded in this
geometry.  As mentioned previously, work in progress indicates that
embedding the horizon in the globally hyperbolic space $H^3$,
following Ref.\ {\cite{ghr09}}, is an elegant way to get around this
restriction.

The cases we examine in detail are associated with
Figs.\ {\ref{fig:embed_a0.0_p6.0_e0.0_thi60}} --
{\ref{fig:embed_a0.85_p4.0_e0.7_thi30}}.  These figures are each a
series of snapshots taken from animations showing the combined orbital
and embedded horizon dynamics.  These animations are available at
{\cite{animations}}.  Readers may find it useful to examine these
visualizations in concert with the text presented below.

\subsection{Embeddings from inclined circular orbits}
\label{sec:inclcirc}

\begin{figure*}[h]
\includegraphics[width = 0.97\textwidth]{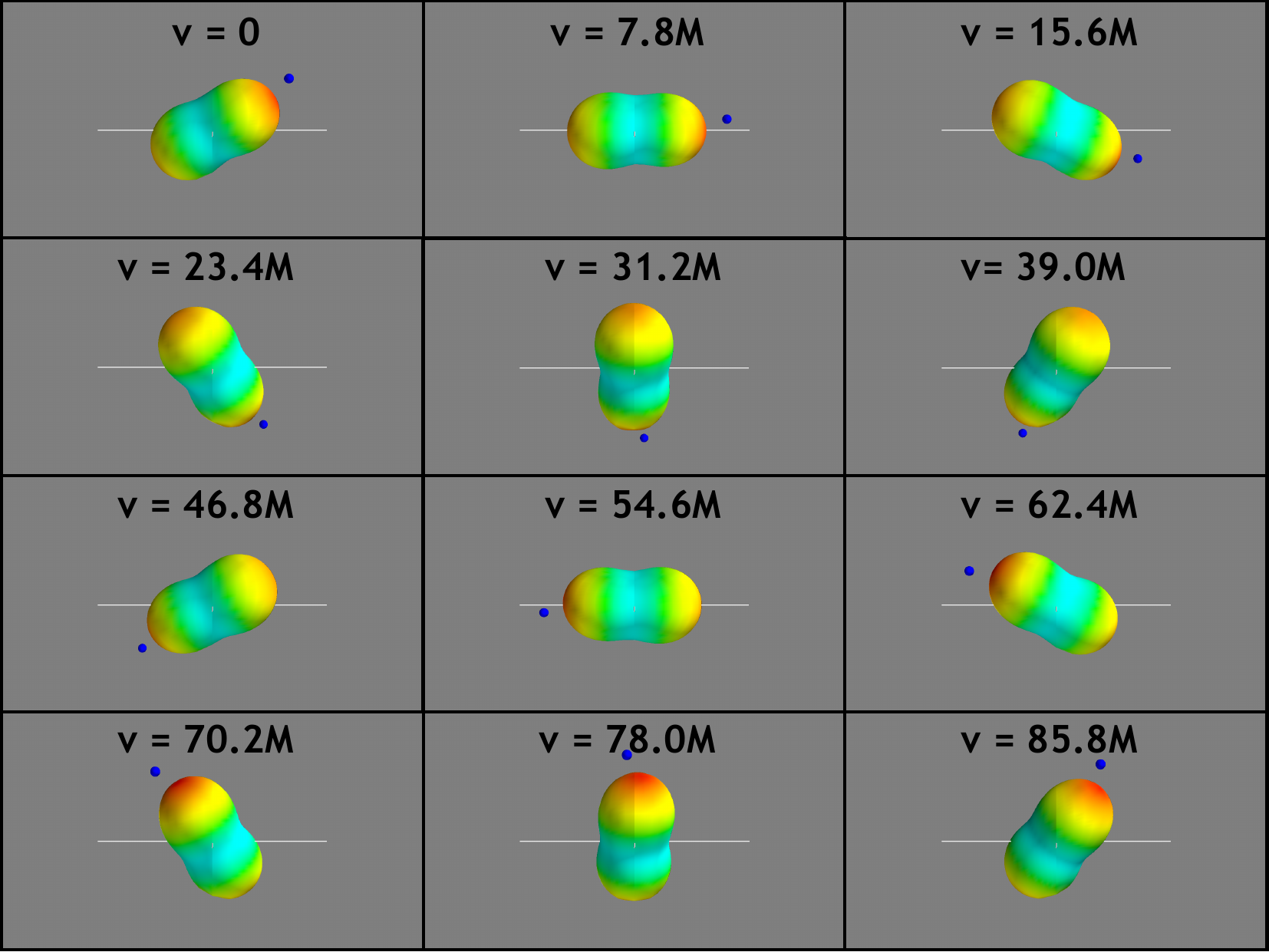}
\vskip 0.3cm
\includegraphics[width = 0.48\textwidth]{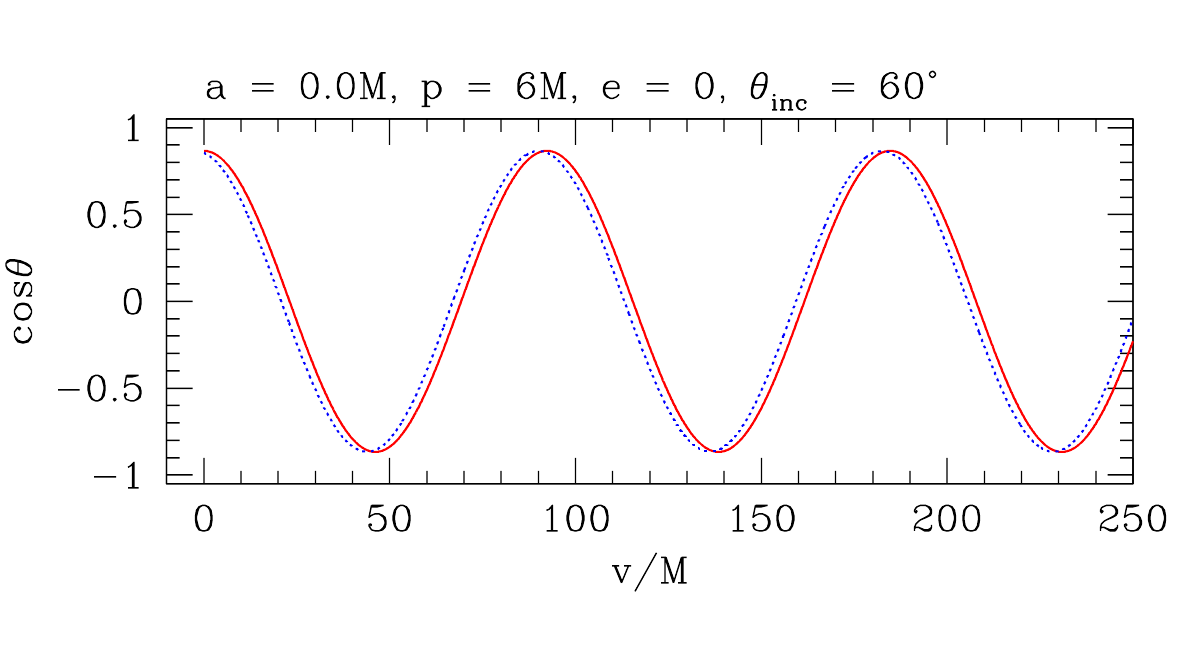}
\includegraphics[width = 0.48\textwidth]{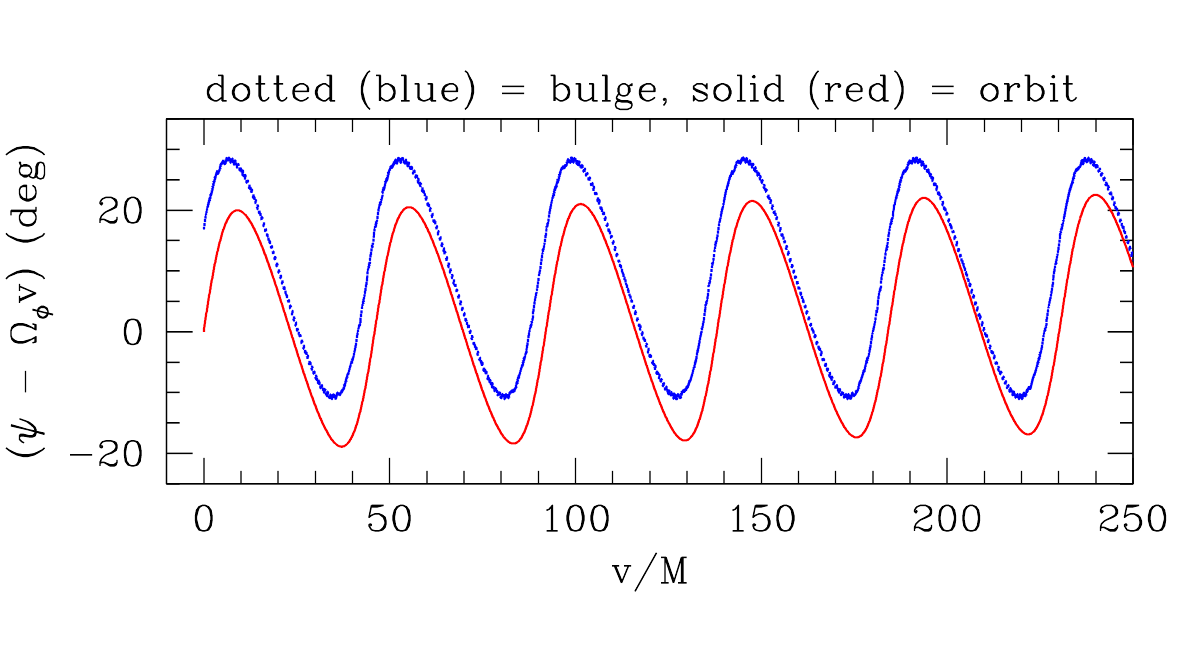}
\vskip -0.3cm
\caption{Top: Snapshots of an animation depicting an embedding of the
  distorted horizon for a circular inclined orbit of a Schwarzschild
  black hole ($a = 0$).  Bottom: Angular position of the horizon's
  bulge [dotted (blue) curve] versus the angular position of the
  orbiting body [solid (red) curve].  The orbit is at radius $r = 6M$,
  inclined at $\theta_{\rm inc} = 60^\circ$ to our chosen equatorial
  plane.  The axes shown in these snapshots indicate the hole's
  equatorial plane; we have placed the camera slightly above this
  plane in order to illustrate the hole's bulge geometry.  The
  orbiting body is indicated by the small ``moon'' (dark blue in color
  plot) located near one of the horizon bulges.  The small body's
  orbit begins on the side of the black hole near the camera, descends
  down through the equatorial plane (crossing just after $v = 7.8M$),
  sweeps behind the far side of the hole (moving from right to left as
  plotted), then comes up through the equatorial plane (crossing just
  after $v = 54.6M$) to pass in front of the side near the camera
  again.  The animation from which these stills are taken is available
  at {\cite{animations}}.  The bottom two panels show the angular
  position of the bulge and the orbit, illustrating the polar angle
  ($\cos\theta$, left) and the axial angle ($\psi$, right).  The
  horizon's bulge moves in lockstep with the orbiting body, always
  leading the orbit by a small, constant angle.}
\label{fig:embed_a0.0_p6.0_e0.0_thi60}
\end{figure*}

We begin with an especially simple case: an inclined, circular orbit
of a Schwarzschild black hole.  Figure
{\ref{fig:embed_a0.0_p6.0_e0.0_thi60}} shows an embedding of the
distorted horizon for the case of a circular orbit with radius $r =
6M$ inclined at $\theta_{\rm inc} = 60^\circ$.  We show 12 frames
illustrating the horizon embedding and particle motion for this orbit;
the frames are evenly spaced over nearly one orbital period ($T_{\rm
  orb} = 92.3M$ for a circular orbit at $r = 6M$ for Schwarzschild).
Axes indicate the location of the equatorial plane; they are static in
this sequence, since the horizon of a Schwarzschild black hole is
static.  The bottom two panels of this figure show the angular
position of the horizon's bulge [defined as the coordinate for which
  the embedding radius is largest; dotted (blue) curves] and the
orbiting body [solid (red) curve], both as functions of ingoing time
$v$.  Bottom left panel shows $\cos\theta(v)$; bottom right shows
$\psi(v) - \Omega_\phi v$.  (We subtract $\Omega_\phi v$ to remove an
uninteresting overall secular growth in $\psi$ over an orbit.)

As should be expected following Sec.\ {\ref{sec:resultsI}}, the
results we see in Fig.\ {\ref{fig:embed_a0.0_p6.0_e0.0_thi60}} are
consistent with the fact that the physics of an inclined orbit is
identical to that of an equatorial orbit in the $a = 0$ limit.  In
particular, the embedded horizon is identical to that shown in the
$r_{\rm orb} = 6M$ panel of paper I's Fig.\ 3, but with the distortion
centered on a plane that is inclined at $\theta_{\rm inc} = 60^\circ$
to our chosen equator.  The offset between the orbit and the horizon's
bulge is constant over the orbit, with the bulge leading the orbiting
body by a fixed amount; this can be seen particularly clearly in an
animation of the horizon and orbit dynamics, and in the plot of
$\cos\theta(v)$.  As we have previously discussed, this can be
understood as due to the spherical symmetry of the Schwarzschild
spacetime --- the magnitude of the tidal field is constant over an
orbit.  In paper I, the lead was purely axial (i.e., purely in the
direction of $\psi$); here it is a mixture of the axial and polar
angles $\psi$ and $\theta$.  As discussed extensively in paper I,
bulge leading orbit is exactly what we expect for circular
Schwarzschild orbits.

\begin{figure*}[h]
\includegraphics[width = 0.97\textwidth]{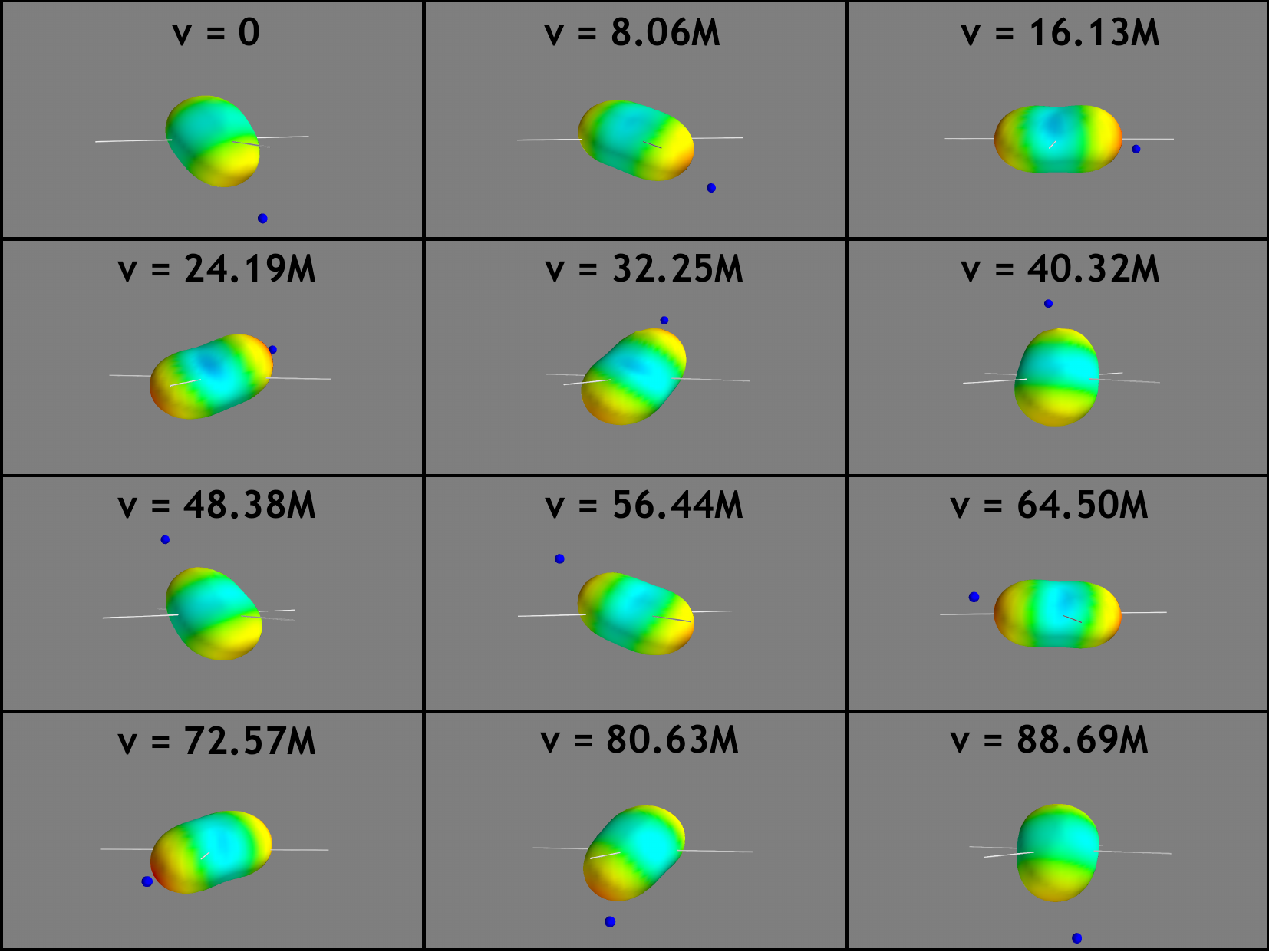}
\vskip 0.3cm
\includegraphics[width = 0.48\textwidth]{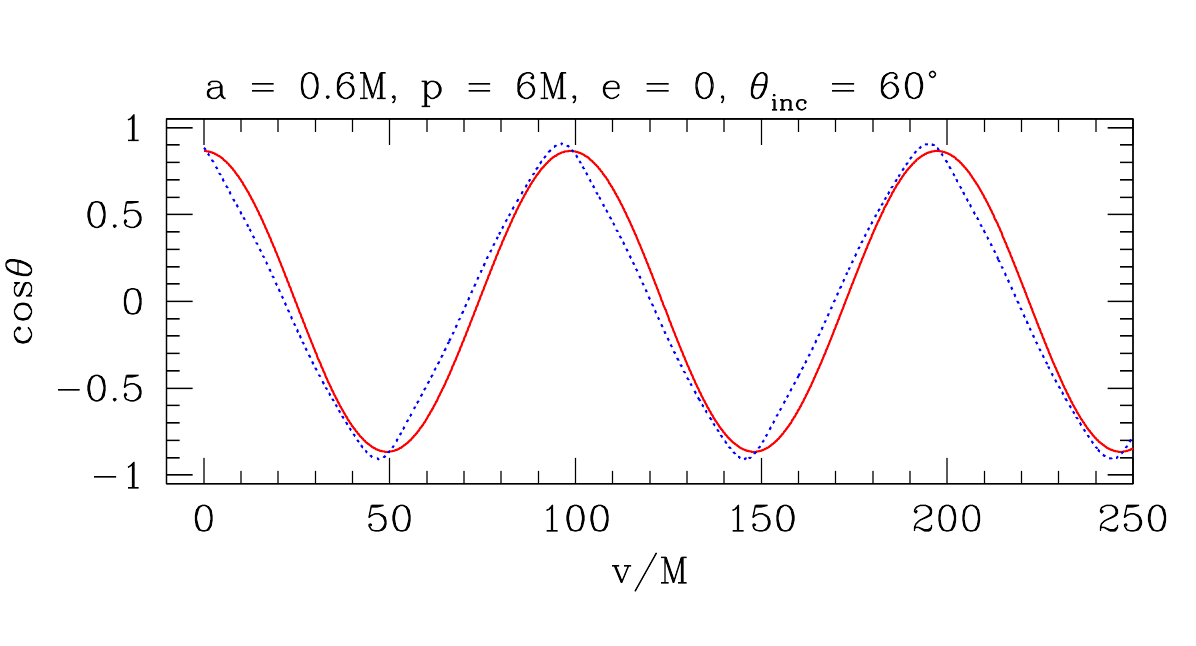}
\includegraphics[width = 0.48\textwidth]{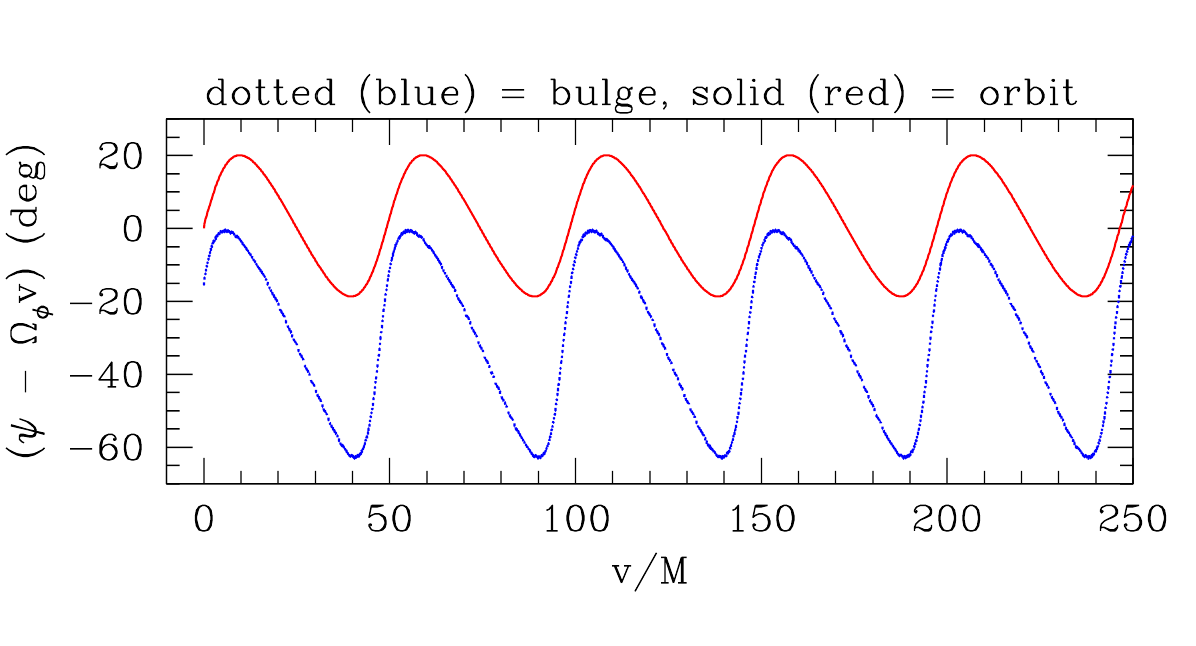}
\vskip -0.3cm
\caption{Identical to Fig.\ {\ref{fig:embed_a0.0_p6.0_e0.0_thi60}},
  except that the black hole shown here is spinning with Kerr
  parameter $a = 0.6M$.  As in
  Fig.\ {\ref{fig:embed_a0.0_p6.0_e0.0_thi60}}, the axes indicate the
  hole's equatorial plane.  In this plot, these axes rotate with the
  event horizon at frequency $\Omega_{\rm H} = a/2Mr_+ = 1/6M$
  (corresponding to a rotation period $T_{\rm H} = 2\pi/\Omega_{\rm H}
  = 37.7M$).  In this sequence, the orbiting body (small sphere, dark
  blue in color) begins near the face close to the camera on the lower
  right-hand side of the black hole.  It then sweeps up, crossing the
  equator soon after $v = 16.13M$, passes behind the black hole, and
  then descends downward again, crossing the equator soon after $v =
  64.5M$.  The horizon's distortions in this case do not move in
  lockstep with the orbiting body.  Instead, the horizon exhibits mild
  shape variations.  This is because the hole is not spherically
  symmetric, and so the tidal field acting on the horizon varies
  slightly over the orbit.  Notice that the horizon's bulge leads the
  angular position of the orbit in the polar ($\theta$) direction, but
  lags its position in the axial ($\psi$) direction; this can be seen
  in the snapshots, but is especially clear in the angle versus time
  plots shown in the bottom two panels.  (We subtract $\Omega_\phi v$
  from $\psi$ to remove the uninteresting secular growth in this angle
  over a single orbit.)  The polar behavior is much like what we see
  for Schwarzschild or very slow rotation; the axial behavior is about
  the same as the behavior we saw for equatorial circular orbits in
  paper I.  The animation from which these stills are taken is
  available at {\cite{animations}}.}
\label{fig:embed_a0.6_p6.0_e0.0_thi60}
\end{figure*}

We next consider an inclined, circular orbit of a Kerr black hole.
Circular Kerr orbits are defined as those for which the
Boyer-Lindquist coordinate radius $r$ is constant.  Although they are
therefore closely tied to a particular coordinate system, they
nonetheless are a well-defined and well-studied subset of Kerr orbits.
It has been shown that the eccentricity $e$ of Kerr orbits [defined in
  Eq.\ (\ref{eq:rofpsi})] decreases over all but the most strong-field
orbits due to gravitational-wave driven backreaction {\cite{gk02,
    th_inprep}}, and that orbits with $e = 0$ remain at $e = 0$
{\cite{ryan96, ko96, mino96}}.  As such, we expect that
gravitational-wave emission will drive large mass-ratio binaries
toward the constant Boyer-Lindquist radius circular limit.

Figure {\ref{fig:embed_a0.6_p6.0_e0.0_thi60}} is much like
Fig.\ {\ref{fig:embed_a0.0_p6.0_e0.0_thi60}}, but for an orbit of a
black hole with spin parameter $a = 0.6M$.  The orbit again has
constant radius $r = 6M$, and is inclined at $\theta_{\rm inc} =
60^\circ$.  We show 12 frames illustrating the horizon and particle
motion for this orbit, with frames evenly spaced over nearly one
orbital period\footnote{``Orbital period'' is somewhat ambiguous for
  this orbit: the period to complete a single polar oscillation is
  $T_{\theta} = 98.7M$, and the period to complete a rotation of
  $2\pi$ radians in the axial direction is $T_\phi = 91.5M$.  These
  two periods differ only by $\sim 10\%$, so our statement that we
  show nearly one period is accurate no matter which notion of period
  we use.}.  In this sequence, the axes (which indicate the equatorial
plane) are tied to the horizon's spin, which completes a full rotation
in a period $T_{\rm H} = 2\pi/\Omega_{\rm H} = 37.7M$.  The bottom two
panels of Fig.\ {\ref{fig:embed_a0.6_p6.0_e0.0_thi60}} compare
$\cos\theta$ and $\psi$ for the horizon's bulge and the orbit's
position.

Some new horizon dynamics begin to appear in
Fig.\ {\ref{fig:embed_a0.6_p6.0_e0.0_thi60}}.  Over the course of an
orbit, the tidal field arising from the small body is not of constant
magnitude since the spacetime is no longer spherically symmetric.  As
a consequence the shape of the embedded horizon varies over an orbit.
There is also interesting new behavior associated with the bulge-orbit
offset.  As discussed at length in paper I (and briefly in
Sec.\ {\ref{sec:intro}}), for circular, equatorial orbits of rapidly
spinning black holes, the horizon bulge tends to lag the position of
the orbiting body on a constant $v$ timeslice.  Let us call this
``Kerr-like'' bulge-orbit behavior, and let us call the opposite
behavior (bulge leading the orbit on a constant $v$ timeslice)
``Schwarzschild-like.''  What we see in
Fig.\ {\ref{fig:embed_a0.6_p6.0_e0.0_thi60}} is that the bulge behaves
in a Kerr-like manner in the $\psi$-direction, but behaves in a
Schwarzschild-like manner in the $\theta$-direction.  This can be seen
by carefully examining the sequence of stills (and the animation from
which these stills are taken), but is especially clear in the bottom
two panels showing the angular position of the orbit and of the
horizon's bulge.

We have found that this bulge-orbit behavior (Schwarzschild-like with
respect to the $\theta$ direction, Kerr-like with respect to the
$\psi$ direction) is quite generic.  It is clear in all the circular,
inclined cases we have examined, and appears in inclined eccentric
examples as well.  This behavior arises from the fact that the black
hole's spin picks out the $\psi$ direction as special.  The hole's
rotation plus the horizon's teleological nature mixes time and axial
angle: a tide that would produce a bulge on a Schwarzschild black hole
at $(\theta_{\rm max},\psi_{\rm max})$ will produce a bulge on a Kerr
black hole at roughly $(\theta_{\rm max}, \psi_{\rm max} -
\delta\psi)$, where $\delta\psi$ is (at leading order) proportional to
the black hole's spin parameter $a$.

\subsection{Horizon embeddings from eccentric orbits}
\label{sec:eccentric}

We conclude our analysis by examining horizon embeddings for highly
eccentric black hole orbits.  The key point to bear in mind here is
that, at leading order, the tidal field varies with orbital separation
as $1/r^3$.  As such, the tidal field from an orbit with eccentricity
$e$ varies by $(1 + e)^3/(1 - e)^3$ over the course of an orbit.  This
factor grows very quickly with $e$.  The two cases we examine in
detail have $e = 0.7$, for which the tide varies by a factor of about
$180$.  This means that the hole can be essentially unaffected by its
companion for much of the orbit, but be highly distorted as the
smaller body passes through periapsis.

\begin{figure*}[h]
\includegraphics[width = 0.97\textwidth]{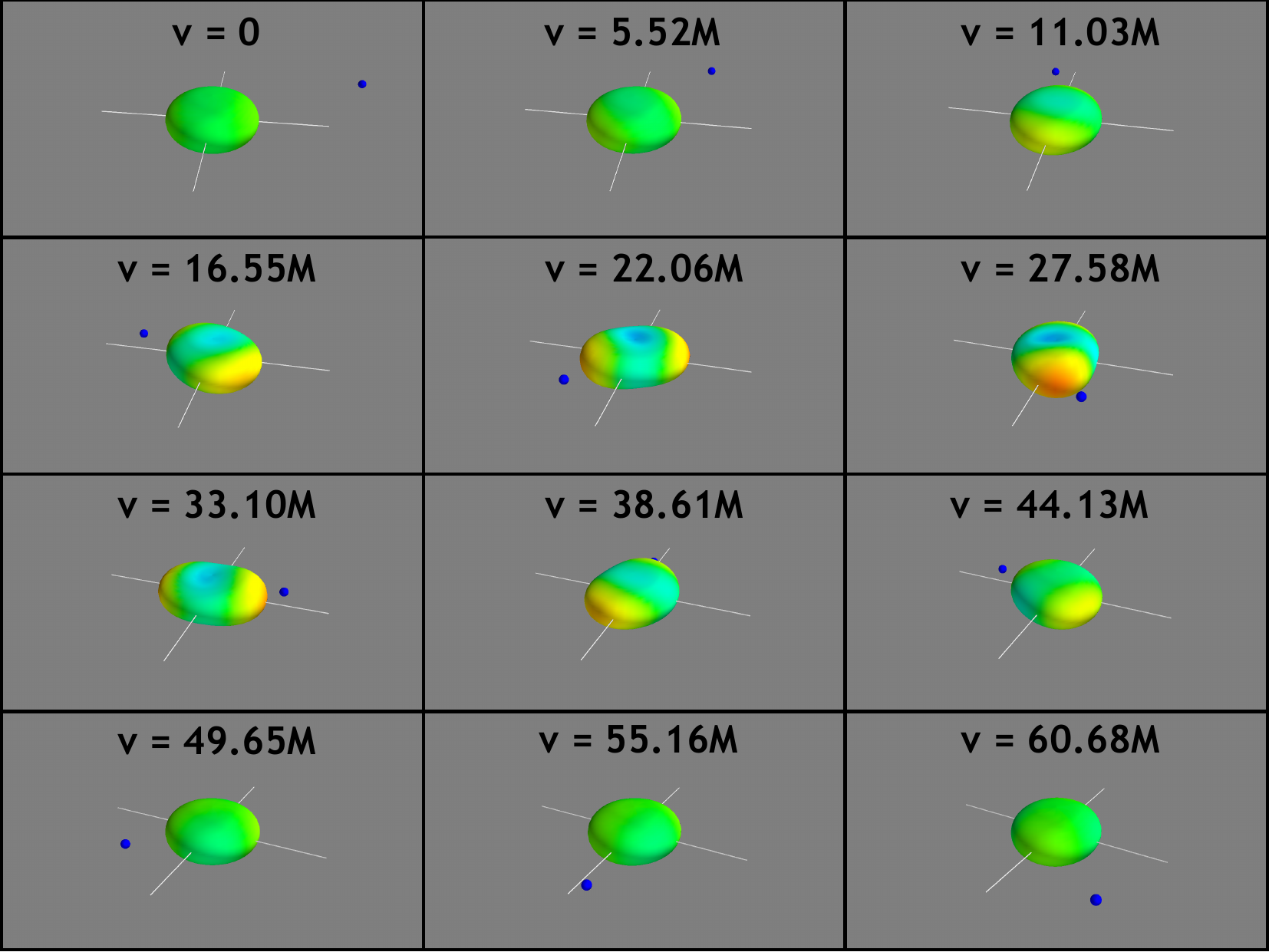}
\vskip 0.3cm
\includegraphics[width = 0.48\textwidth]{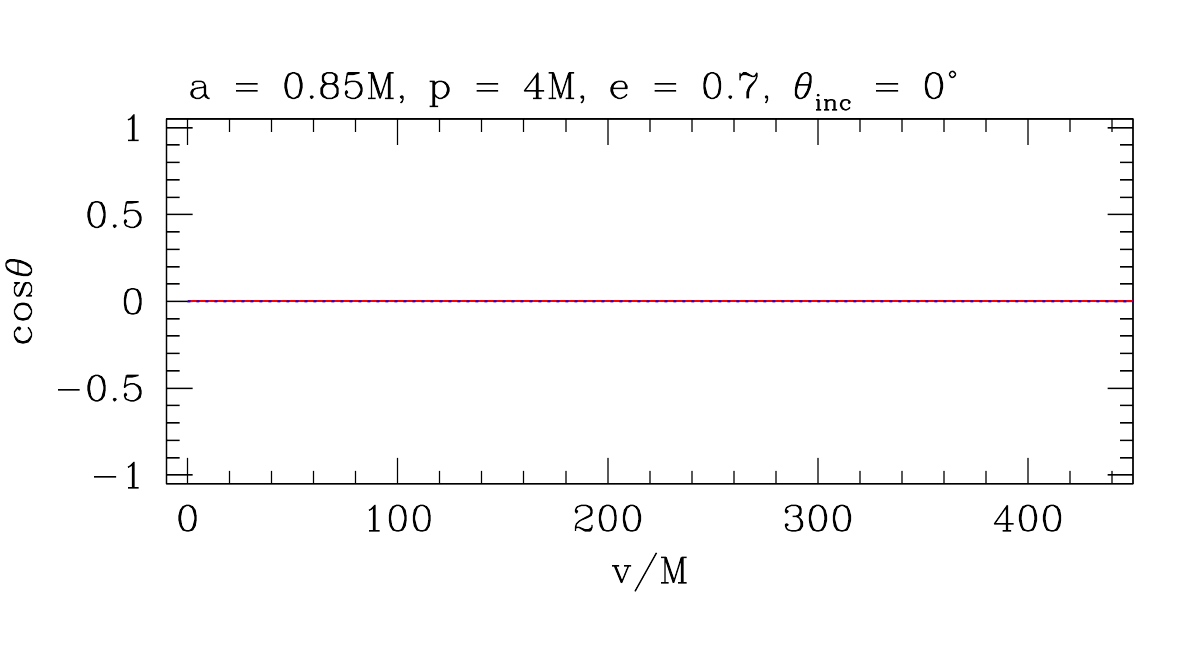}
\includegraphics[width = 0.48\textwidth]{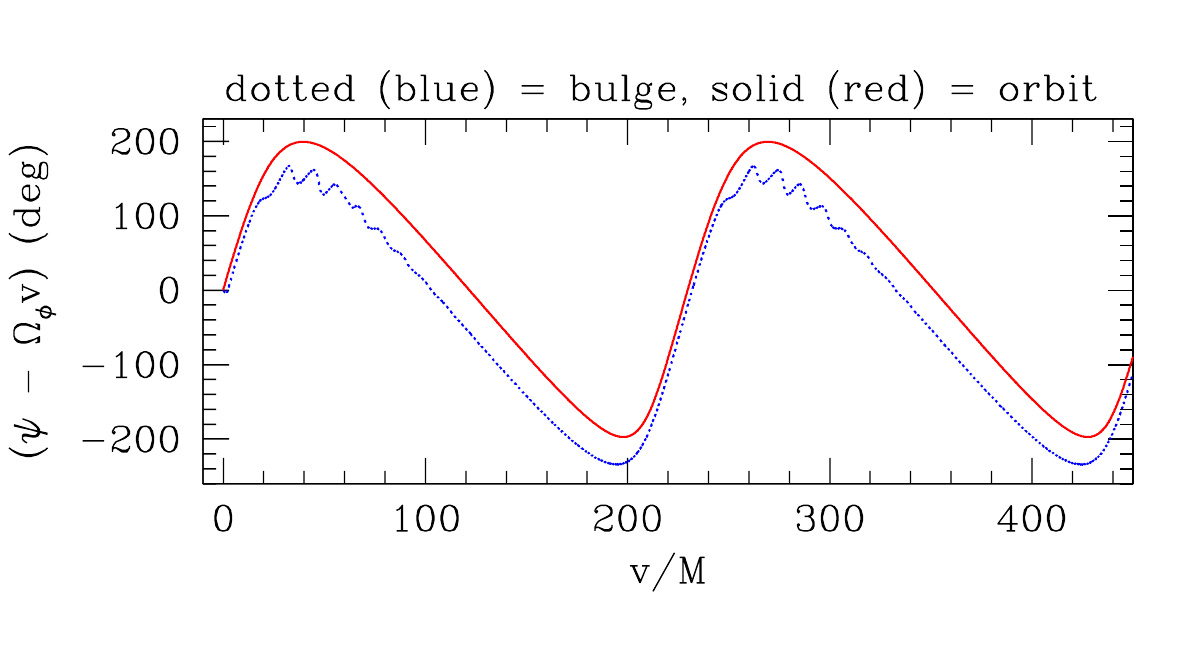}
\vskip -0.3cm
\caption{Snapshots of an animation depicting an embedding of the
  distorted horizon for an equatorial eccentric orbit of a rapidly
  spinning Kerr black hole ($a = 0.85M$).  The spin is nearly the
  largest value for which a Euclidean embedding of the undistorted
  horizon exists; the undistorted embedding geometry at this spin is
  of an axially symmetric oblate spheroid that is nearly flat at its
  poles.  The orbit has semi-latus rectum $p = 4M$ and eccentricity $e
  = 0.7$, so its orbital radius varies from $r_{\rm max} = 13.33M$ to
  $r_{\rm min} = 2.34M$.  As in
  Fig.\ {\ref{fig:embed_a0.6_p6.0_e0.0_thi60}}, the axes indicating
  the equatorial plane rotate at the horizon frequency $\Omega_{\rm H}
  = a/2Mr_+$, corresponding to period $T_{\rm H} = 22.6M$.  The axes
  complete a quarter turn every $5.6M$; this is very close to the
  cadence with which we sample this animation, so the axes appear
  nearly stationary in this sequence.  The orbiting body is shown
  moving from roughly $r_{\rm max}/2$ to $r_{\rm min}$, and back out
  to roughly $r_{\rm max}/2$.  For the first and last few stills shown
  here, the embedded horizon is nearly identical to that of an
  undistorted Kerr black hole.  The embedded horizon by contrast is
  highly distorted in stills corresponding to $r \simeq r_{\rm min}$
  ($16.55 M \le v \le 44.13M$).  This reflects the fact that the tidal
  field varies at leading order as $1/r^3$, which changes by a factor
  $(1 + e)^3/(1 - e)^3$ over an eccentric orbit.  The tidal field thus
  varies by a factor $\sim 180$ for this orbit; even over this limited
  segment (for which the orbit only goes out to about $r_{\rm
    max}/2$), the tidal field varies by $\sim 180/8 \simeq 22$.  The
  horizon bulge lags the particle's position in $\psi$ at all times,
  consistent with the behavior seen and discussed in paper I for
  rapidly rotating Kerr black holes.  (The $\theta$ behavior is
  uninteresting, since this is an equtorial orbit.)  The
  high-frequency wiggles in the $\psi$-position of the bulge near $v
  \simeq 40M$ and $v \simeq 260M$ are perhaps related to the high-spin
  phenomenon discussed in Sec.\ {\ref{sec:highspinwiggles}}.  The
  animation from which these stills are taken is available at
  {\cite{animations}}.}
\label{fig:embed_a0.85_p4.0_e0.7_thi0}
\end{figure*}

Figure {\ref{fig:embed_a0.85_p4.0_e0.7_thi0}} shows this behavior
quite clearly; see {\cite{animations}} for the animation from which
these stills were taken.  The large black hole used here has spin $a =
0.85M$, nearly the largest value for which a globally Euclidean
embedding exists.  The orbit is equatorial ($\theta_{\rm inc} =
0^\circ$), quite strong field ($p = 4M$), and highly eccentric ($e =
0.7$).  We only show a portion of a full radial cycle, from $r \simeq
r_{\rm max}/2$ to $r_{\rm min} = p/(1 + e)$ back to $r \simeq r_{\rm
  max}/2$.  As in Fig.\ {\ref{fig:embed_a0.6_p6.0_e0.0_thi60}}, the
axes indicating the equatorial plane rotate with the horizon.  For $a
= 0.85M$, the period of this rotation is $T_{\rm H} = 22.6M$.  We
sample our animation every $5.6M$.  By coincidence, this is nearly
$T_{\rm H}/4$, so the axes are sampled in a nearly stroboscopic
fashion, and appear to be stationary.

The embedded horizon of an undistorted $a = 0.85M$ black hole is an
oblate ellipsoid that is nearly flat at the poles.  This geometry can
be seen in the first and last few frames shown in
Fig.\ {\ref{fig:embed_a0.85_p4.0_e0.7_thi0}} --- the tidal field is so
weak in these frames\footnote{This is why we show only a fraction of
  an orbit here.  A full radial cycle of this orbit takes $T_r =
  229.8M$, with the tide having a large impact only for $r \simeq
  r_{\rm min}$.  The hole is practically undistorted for the majority
  of the orbit.}  (for which $r \sim r_{\rm max}/2$) that the horizon
is not noticeably distorted by the companion.  The distortion becomes
quite strong as the orbital approaches periapsis: we see the horizon
beginning to change shape at $v = 11.03M$, and is highly distorted
over the range $22.06M \le v \le 33.10M$.  At its peak, the horizon's
distortion is similar to the most distorted horizon embedding shown in
paper I, the right-hand panel of that paper's Fig.\ 7.  Notice the
Kerr-like bulge-orbit behavior: the bulge's position in $\psi$ lags
the orbit in all cases.  This is quite clear in the $v = 27.58M$
panel, and in the plot of $\psi(v)$.  (Since the orbit is equatorial,
there is no lag or lead associated with $\theta$.)

\begin{figure*}[h]
\includegraphics[width = 0.97\textwidth]{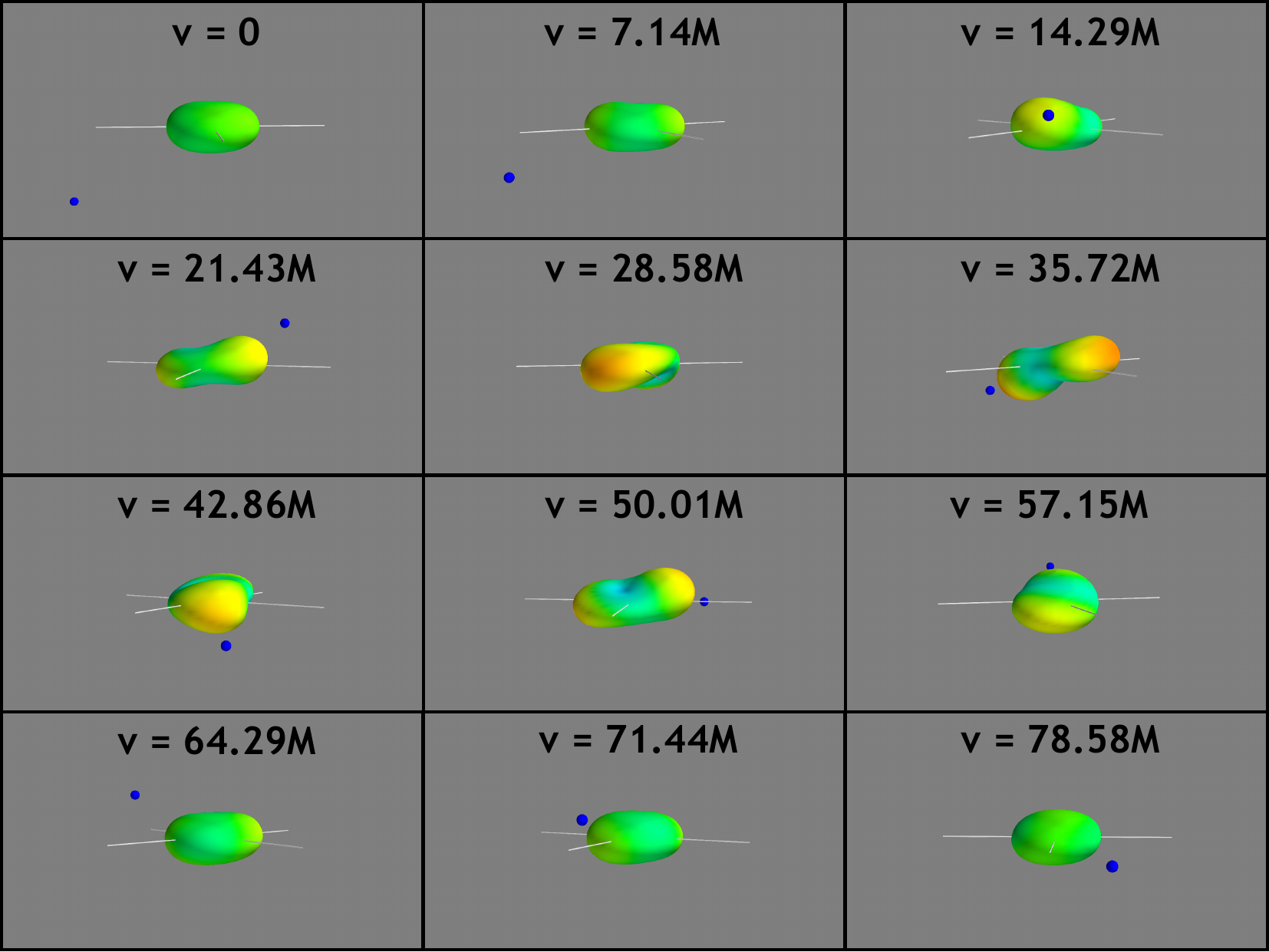}
\vskip 0.3cm
\includegraphics[width = 0.48\textwidth]{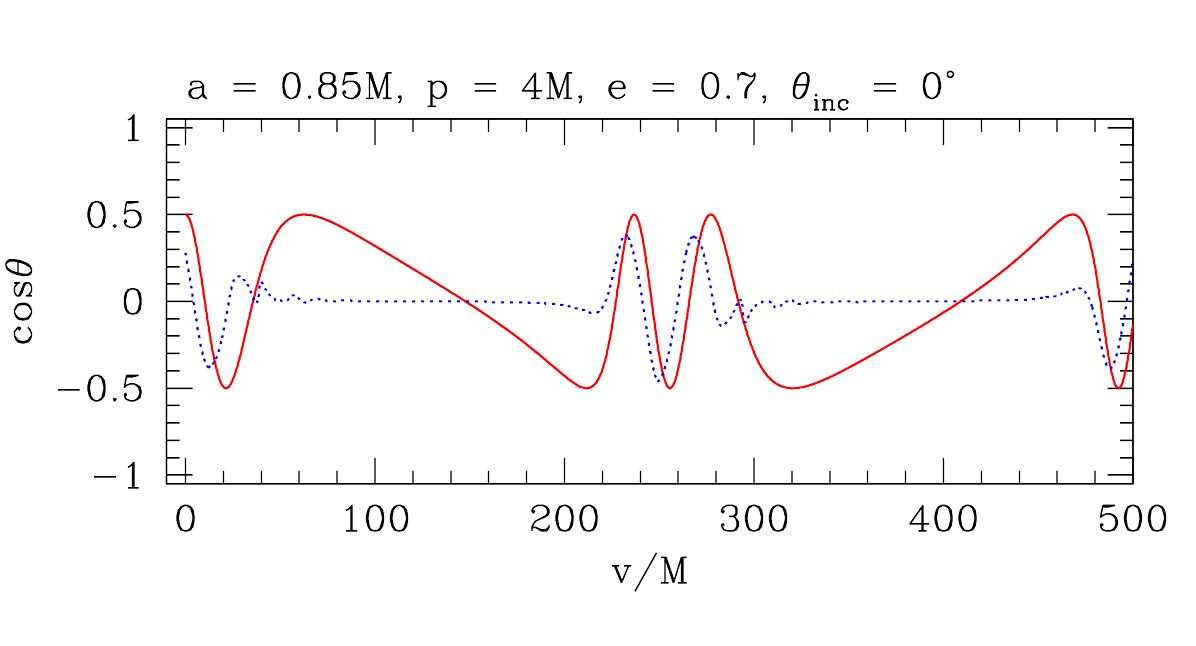}
\includegraphics[width = 0.48\textwidth]{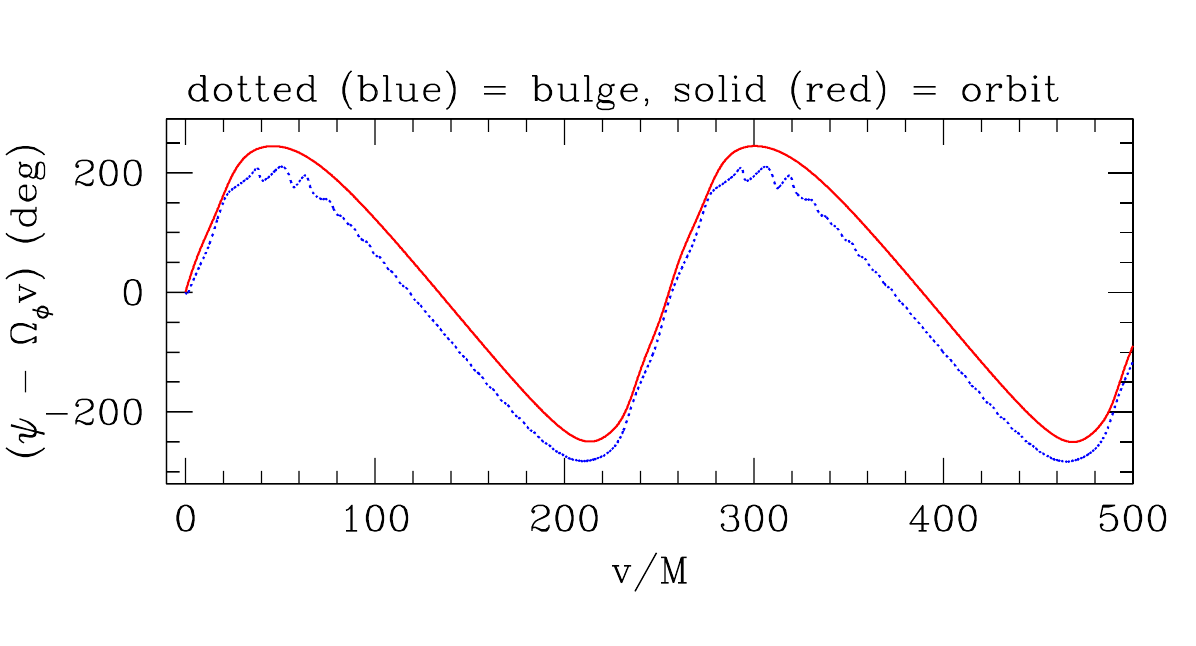}
\vskip -0.3cm
\caption{Snapshots of an animation depicting an embedding of the
  distorted horizon for a generic orbit of a rapidly spinning Kerr
  black hole ($a = 0.85M$).  The system is nearly identical to that
  used in Fig.\ {\ref{fig:embed_a0.85_p4.0_e0.7_thi0}}, but we have
  inclined the orbit to $\theta_{\rm inc} = 30^\circ$.  The orbiting
  body is again shown moving from roughly $r_{\rm max}/2$ to $r_{\rm
    min}$, and back out to roughly $r_{\rm max}/2$.  The horizon's
  dynamics here shares features with both the equatorial case depicted
  in Fig.\ {\ref{fig:embed_a0.85_p4.0_e0.7_thi0}} and the inclined
  cases in Figs.\ {\ref{fig:embed_a0.0_p6.0_e0.0_thi60}} and
  {\ref{fig:embed_a0.6_p6.0_e0.0_thi60}}.  In particular, the horizon
  varies from nearly undistorted when $r \simeq r_{\rm max}/2$
  (roughly first and last stills in this sequence) to highly distorted
  when $r \simeq r_{\rm min}$ (stills from $28.52M \le v \le 57.15M$),
  in a manner qualitatively similar to the eccentric equatorial case.
  However, the horizon bulge flexes above and below the plane as the
  orbital motion oscillates in the polar direction, very much like the
  circular inclined cases.  The panels illustrating $\cos\theta$ and
  $\psi$ versus time shows that the bulge lags the body in $\psi$.  At
  periapsis, the bulge lags the body the body in $\theta$.  This is
  exactly the same offset behavior that was seen in the circular
  inclined Kerr case shown in
  Fig.\ {\ref{fig:embed_a0.6_p6.0_e0.0_thi60}}.  (When the orbiting
  body is far from periapsis, the location of the bulge is difficult
  to determine accurately; our code returns $\theta = 90^\circ$ for
  the bulge's position, reflecting the oblate spheroid shape of the
  undistorted black hole.)  The animation from which these stills are
  taken is available at {\cite{animations}}.}
\label{fig:embed_a0.85_p4.0_e0.7_thi30}
\end{figure*}

Notice also the high-frequency, low-amplitude wiggles in the $\psi$
position of the horizon's bulge at $v \simeq 40M$ and $v \simeq 260M$.
These are reminiscent of the high-spin features that we discussed in
Sec.\ {\ref{sec:highspinwiggles}}.  In this case, we do not see such
strong wiggles in the shear $\sigma$.  It is plausible that the
wiggles are present in $\sigma$, but at such low amplitude that they
cannot be cleanly pulled out of that data; it could be that
constructing other quantities associated with the horizon distortion,
such as $R^{(1)}_{\rm H}$ and the embedding surface, makes the wiggles
stand out even more strongly.  Similar behavior is seen near periapsis
for the generic case we discuss next (cf.\ lower right-hand panel of
Fig.\ {\ref{fig:embed_a0.85_p4.0_e0.7_thi30}}).  We hope to study this
further in future work.

Figure {\ref{fig:embed_a0.85_p4.0_e0.7_thi30}} shows the embedding for
a horizon distorted by tides from a generic orbit.  We again consider
spin $a = 0.85M$, and a very strong-field ($p = 4M$), highly eccentric
($e = 0.7$) orbit, but we now take the orbit to be inclined at
$\theta_{\rm inc} = 30^\circ$.  The set of frames we show again
corresponds to motion from roughly $r_{\rm max}/2$ to $r_{\rm min}$
and back to nearly $r_{\rm max}/2$.  We have moved the ``camera'' in
this sequence to a point slightly above the equatorial plane in order
to more clearly see the orbit's polar motion, and the distortions
associated with motion above and below this plane.

The embedding dynamics shown in
Fig.\ {\ref{fig:embed_a0.85_p4.0_e0.7_thi30}} combines the features
found for inclined circular orbits with those found for eccentric
equatorial orbits.  In particular, notice that the embedded horizon
geometry is practically undistorted in the first frame, as well as the
last two or so frames.  This again reflects the large range of the
tidal field that acts on the horizon for eccentric orbits; when $r
\gtrsim r_{\rm max}/2$, the horizon's distortions are so mild that
they cannot be seen in these graphics.  A full radial cycle of this
orbit takes $T_r = 255.1M$, so the horizon is practically undistorted
for a large fraction of this orbit.  As the orbit oscillates above and
below the equatorial plane, the horizon's bulge likewise oscillates
above and below the plane.  The bulge lags the orbit's $\psi$
position, but leads\footnote{The $\theta$ behavior of the bulge is
  only clear when the orbit is at periapsis.  When the orbit is far
  from periapsis, the tidal deformation is gentle, and our algorithm
  for determining the position of the bulge becomes inaccurate due to
  discretization errors.  The algorithm returns $\theta_{\rm bulge} =
  90^\circ$ in this case, corresponding to the largest radius of the
  undistorted Kerr embedding.} its $\theta$ position.  This is
basically the same behavior that we saw for the inclined circular Kerr
orbit (Fig.\ {\ref{fig:embed_a0.6_p6.0_e0.0_thi60}}) --- Kerr-like in
the axial direction, Schwarzschild-like in the polar direction.
Having already examined the equatorial and the circular limits in
detail, there are no surprises in
Fig.\ {\ref{fig:embed_a0.85_p4.0_e0.7_thi30}}.  The interesting
behaviors seen in the previously considered cases combine in the
generic case in a very logical way.

\section{Conclusions}

In this paper, we have taken the tools that we introduced in paper I
for studying event horizons that are distorted by a strong-field (but
small mass ratio) binary companion, and have applied them to eccentric
and inclined binaries.  For such orbits, the on-horizon tidal field
varies significantly over the course of an orbit, leading to dynamical
event horizon behavior.  We have studied these horizon dynamics with
multiple measures, examining the phase offset between the applied tide
and the resulting shear to the horizons, as well as examining
embeddings of the distorted horizons in a globally Euclidean 3-space.

Many of the results we have found follow in a fairly natural and
logical way from results that were shown in paper I.  In particular,
we find that tidal bulges tend to lead the position of the orbiting
body for very slow black hole spin, but lag the orbit for fast black
hole spin.  This is exactly the teleological tidal behavior that was
seen with the simpler orbits we examined in paper I.  We find an
interesting variant of this behavior in the present analysis by
looking at orbits that are inclined with respect to the hole's
equatorial plane: the bulge tends to lead the orbit in the $\theta$
direction (``Schwarzschild-like'' behavior), but lags the orbit for
rapid spin in the $\psi$ direction (``Kerr-like'' behavior).  The fact
that the bulge exhibits different behavior with respect to the two
angles is not surprising, since the hole rotates in the direction of
$\psi$.

One interesting new behavior we have found are the low-amplitude,
high-frequency wiggles which appear in the shear $\sigma$ associated
with the distortion of nearly extremal ($a \gtrsim 0.9995M$) black
holes.  A perhaps related low-amplitude, high-frequency wiggle is
apparent in the horizon embedding of more slowly rotating ($a =
0.85M$) black holes.  We have not succeeded in constructing a
compelling explanation for these features.  Although we can estimate
the frequency of the wiggles fairly well, we cannot link them to other
frequencies in the problem; and, the rate at which the oscillations
decay with time does not appear to relate to other timescales in the
problem, such as the correlation time $\kappa^{-1}$ associated with
the Green's function which connects to the tide to the shear.  We hope
that future work will elucidate the nature of this interesting
phenomenon.

\acknowledgments

We are grateful to Eric Poisson for many useful discussions regarding
tidally distorted black holes, and to Daniel Kennefick for discussions
regarding the connection between tidal coupling and superradiant
Teukolsky equation modes, which did much to inspire this analysis.  We
also thank this paper's referee, whose criticisms led us to expand
much of our discussion, and to significantly re-examine some of the
claims and analyses we presented in a previous version of this paper.
Our work on this problem has been supported by NSF grant PHY-1403261.

\appendix

\section{Newman-Penrose fields and on-horizon tensors}
\label{app:onhorizmap}

In this paper, we work with quantities that are based on
Newman-Penrose fields such as the complex curvature scalar $\Psi_0$.
Other papers, notably VPM11, use tensors which live in the manifold
defined by the black hole's event horizon.  There is a simple
one-to-one correspondence between these two representations for the
quantities which are important for our analysis.  We develop this
correspondence in this appendix.

We begin by defining some notation and background.  As elsewhere in
this paper, we use ingoing Kerr coordinates $(v,r,\theta,\psi)$ here.
Components of tensors in the 2-dimensional manifold of the black
hole's event horizon are labeled with upper-case Latin indices; these
components range over the set $(\theta,\psi)$.  (As elsewhere, Greek
indices denote tensors in 4-dimensional spacetime.)  Define the
projection tensor ${P^A}_\alpha$, whose components in ingoing Kerr
coordinates are given by the matrix
\begin{equation}
{P^A}_\alpha \doteq
\begin{pmatrix}
0 & 0 & 1 & 0\cr
0 & 0 & 0 & 1
\end{pmatrix}\;;
\end{equation}
the components of the inverse tensor ${P^\beta}_B$ are the transpose
of this.  When this operates on tensors at $r = r_+$, it projects
quantities onto a slice of constant $v$ on the horizon.  At a given
moment on the horizon, the spacetime's line element
(\ref{eq:Kerr_ingoing}) becomes
\begin{equation}
ds^2 = g_{AB} dx^A dx^B = \Sigma_+\,d\theta^2 +
\frac{4M^2r_+^2\sin^2\theta}{\Sigma_+}d\psi^2\;,
\end{equation}
where $\Sigma_+ = r_+^2 + a^2\cos^2\theta$.  Finally, we will need the
Newman-Penrose null legs in the Hawking-Hartle representation
{\cite{hh72,tp74}}; these are given in
Eqs.\ (\ref{eq:lHH})--(\ref{eq:mHH}).  For $r \to r_+$,
\begin{eqnarray}
l^\mu &\to& \left[1,0,0,a\Omega_{\rm H}\right]\;,
\\
m^\mu &\to& \frac{1}{\sqrt{2}(r_+ + ia\cos\theta)}\left[0, 0, 1,
    i\left(\csc\theta - a\Omega_{\rm H}\sin\theta\right)\right]\;.
\nonumber\\
\end{eqnarray}
(We will not need $n^\mu$.)  At $r = r_+$, $l^\mu$ is tangent to the
null generators of an unperturbed Kerr hole's horizon.  Let us
manipulate $m^\mu(r_+)$: we write
\begin{eqnarray}
m^A(r_+) &\equiv& m^\mu(r_+) {P^A}_\mu
\nonumber\\
&=& \frac{1}{\sqrt{2}}\left(\alpha^A + i\beta^A\right)\;,
\label{eq:mleg_rewrite}
\end{eqnarray}
where
\begin{eqnarray}
\alpha^A &\doteq& \frac{1}{\Sigma_+}\left[r_+,
a\cos\theta\left(\csc\theta - a\Omega_{\rm H}\sin\theta\right)\right]\;,
\label{eq:alphadef}
\\
\beta^A &\doteq& \frac{1}{\Sigma_+}\left[-a\cos\theta,
r_+\left(\csc\theta - a\Omega_{\rm H}\sin\theta\right)\right]\;.
\label{eq:betadef}
\end{eqnarray}
Notice that $g_{AB}\alpha^A\alpha^B = g_{AB}\beta^A\beta^B = 1$,
$g_{AB}\alpha^A\beta^B = 0$.

The intrinsic geometry of the horizon is governed by the Weyl
curvature.  In our analysis, we use the Newman-Penrose scalar
$\Psi_0$, which is given by
\begin{equation}
\Psi_0 = -C_{\mu\alpha\nu\beta}l^\mu m^\alpha l^\nu m^\beta\;.
\label{eq:psi0def2}
\end{equation}
Our focus is on this quantity on the horizon.  Let us define
\begin{equation}
C_{AB} \equiv \left(C_{\mu\alpha\nu\beta}\,l^\mu {P^\alpha}_A\, l^\nu
  {P^\beta}_B\right)_{r_+}\;,
\end{equation}
where the subscripted $r_+$ means that all the quantities in
parentheses are to be evaluated at $r = r_+$.  This definition is
identical to that in VPM11 [see text following their Eq.\ (2.30)].
Using this, on the horizon we have
\begin{eqnarray}
\Psi_0 &=& -C_{AB} m^A m^B
\nonumber\\
&=& -\frac{1}{2}C_{AB}\left(\alpha^A\alpha^B - \beta^A\beta^B +
i\alpha^A\beta^B + i\beta^A\alpha^B\right)
\nonumber\\
&\equiv& -C_{AB}\left(\boldsymbol{e}_+^{AB} + i
\boldsymbol{e}_\times^{AB}\right)\;.
\label{eq:psi0_1}
\end{eqnarray}
On the second line, we used Eq.\ (\ref{eq:mleg_rewrite}); on the
third, we introduced the polarization tensors
\begin{eqnarray}
\boldsymbol{e}_+^{AB} &=& \frac{1}{2}\left(\alpha^A \alpha^B - \beta^A
\beta^B\right)\;,
\\ 
\boldsymbol{e}_\times^{AB} &=& \frac{1}{2}\left(\alpha^A \beta^B +
\beta^A \alpha^B\right)\;.
\end{eqnarray}
We further simplify Eq.\ (\ref{eq:psi0_1}) by defining the Weyl
polarization components:
\begin{eqnarray}
C_+ &\equiv& C_{AB} \boldsymbol{e}_+^{AB}\;,
\\
C_\times &\equiv& C_{AB} \boldsymbol{e}_\times^{AB}\;,
\end{eqnarray}
yielding
\begin{equation}
\Psi_0 = -\left(C_+ + iC_\times\right)\;.
\end{equation}
In other words, the on-horizon Weyl polarizations are simply the real
and imaginary parts of $\Psi_0$ on the horizon, modulo an overall
sign.

Lowering indices on the polarization tensors,
\begin{eqnarray}
\boldsymbol{e}^+_{AB} &=& g_{AC} g_{BD} \boldsymbol{e}_+^{CD}\;,
\label{eq:eplus_def}\\
\boldsymbol{e}^\times_{AB} &=& g_{AC} g_{BD} \boldsymbol{e}_\times^{CD}\;,
\label{eq:ecross_def}
\end{eqnarray}
allows us to construct the on-horizon Weyl tensor from the
polarization components:
\begin{equation}
C_{AB} = C_+ \boldsymbol{e}^+_{AB} + C_\times \boldsymbol{e}^\times_{AB}\;.
\end{equation}

Another important Newman-Penrose quantity which we can analyze in this
manner is the spin coefficient
\begin{equation}
\sigma = m^\mu m^\nu \nabla_\mu l_\nu\;.
\end{equation}
At $r = r_+$, this describes the shear of the horizon's generators.
Because $m^v = m^r = 0$ at $r = r_+$, we have
\begin{equation}
\sigma(r_+) = \left(m^A m^B \nabla_A l_B\right)_{r_+}\;.
\end{equation}
Define
\begin{equation}
\sigma_{AB} = \frac{1}{2}\left(\nabla_A l_B + \nabla_B l_A\right)_{r_+}\;.
\end{equation}
Note that $\sigma_{AB}$ is trace free since $\nabla_A l^A = 0$ on the
horizon.  This definition of the shear tensor for the horizon's null
generators is therefore equivalent to that used in VPM11 [compare
  their Eqs.\ (2.11) and (2.15)].  Using
Eqs.\ (\ref{eq:mleg_rewrite}), (\ref{eq:eplus_def}), and
(\ref{eq:ecross_def}), we find
\begin{eqnarray}
\sigma(r_+) &=& \sigma_{AB}\left(\boldsymbol{e}_+^{AB} + i
\boldsymbol{e}_\times^{AB}\right)
\nonumber\\
&=& \sigma_+ + i \sigma_\times\;.
\end{eqnarray}
The shear polarizations written here were introduced by VPM11; they
are defined in a manner analogous to $C_+$ and $C_\times$, and are
just the real and the imaginary parts of the Newman-Penrose quantity
$\sigma$.

\end{document}